\documentclass[reprint, prd, twocolumn,floatfix]{revtex4}
\usepackage{color}
\usepackage{hyperref}
\usepackage[english]{babel}
\usepackage{inputenc}
\usepackage{enumerate}
\usepackage{amsfonts}
\usepackage{amssymb}
\usepackage{amsmath}
\usepackage{dcolumn}
\usepackage{bm}
\usepackage{graphicx, graphics}
\usepackage{graphicx}
\usepackage{natbib}
\begin{document}
\title{Heavy quarkonia spectroscopy at zero and finite temperature in bottom-up AdS/QCD}

\author{Miguel Angel Martin Contreras}
\email{miguelangel.martin@uv.cl}
\affiliation{%
 Instituto de F\'isica y Astronom\'ia, \\
 Universidad de Valpara\'iso,\\
 A. Gran Breta\~na 1111, Valpara\'iso, Chile
}

\author{Alfredo Vega}%
 \email{alfredo.vega@uv.cl}
\affiliation{%
 Instituto de F\'isica y Astronom\'ia, \\
 Universidad de Valpara\'iso,\\
 A. Gran Breta\~na 1111, Valpara\'iso, Chile
}

\author{Saulo Diles}
\email{smdiles@ufpa.br}
\affiliation{Campus Salin\'opolis,\\ Universidade Federal do Par\'a,\\
68721-000, Salin\'opolis, Par\'a, Brazil}

\begin{abstract}
S-wave states of charmonium and bottomonium are described using bottom-up AdS/QCD. We propose a holographic model that unifies the description of masses and decay constants, leading to a precise match with experimental data on heavy quarkonia. Finite temperature effects are considered by calculating the current-current spectral functions of heavy vector mesons.
The identification of quasi-particle states as Breit-Wigner resonances in the holographic spectral function was made.
We develop a prescription to subtract background contributions from the spectral function to isolate the Breit-Wigner peak. The quasi-particle holographic thermal evolution is described, allowing us to estimate the melting temperature for vector charmonia and bottomonia. Our holographic model predicts that $J/\Psi$ melts at $415$ MeV $(\sim 2.92 ~T_c)$   and $\Upsilon$ melts at $465$ MeV $(\sim 3.27~ T_c)$.
\end{abstract}
  % keywords can be removed

\maketitle

\section{Introduction}\label{intro}
Heavy quarkonia work as a probe of quark-gluon plasma formation in heavy-ion collisions, where  charmonium suppression seemed to play the fundamental role \cite{Matsui:1986dk}. It happens that $J/\Psi$ track is hard to reconstruct due to physical effects such as nuclear absorption and recombination \cite{Chaudhuri:2002df,Liu:2010ej,Abreu:2017cof}. This difficulty in tracking back the charmonium trajectories made unfavorable $J/\Psi$ as a precise probe of QGP.  On the other hand, bottomonium production by recombination and regeneration effects is small \cite{Song:2011nu, Emerick:2011xu,Reed:2011fr}.  Bottomonium then emerges as a promising candidate to probe QGP properties, but not invalidating the importance of charmonium in this context. See \cite{Krouppa:2018lkt, Yao:2018sgn}.
 
Charmonium and bottomonium mesons were experimentally discovered, latter a than its light cousins ($\rho,\phi$), due to its considerable threshold energies imposed by the heavy $c,b$ quark masses. Curiously, current experimental data about the spectrum of radial excitations is more extensive and complete for the heavy vector than the light ones. The decay constants for the excited S-wave states are entirely determined from experiments for the heavy vector quarkonium \cite{Tanabashi:2018oca}. Decay constants of charmonium and bottomonium are observed to be decreasing with excitation levels. For the $\phi$ meson, the decay constants of excited states are estimated from experimental data. These estimations predict they are also decreasing with excitation level \cite{Pang:2019ttv,Badalian:2019xir}.
 
 Meson spectroscopy is a static low energy phenomenon. In this case, the color interaction is strongly coupled and a non-perturbative approach for strong interactions is required  \cite{Gross:1973id,Politzer:1973fx,vanRitbergen:1997va}. One important non-perturbative approach is the holographic dual of QCD, referred as AdS/QCD \cite{Polchinski:2001tt, BoschiFilho:2002vd,Erlich:2005qh, Brodsky:2007hb}. AdS/QCD models are inspired in the exact duality between the conformal and supersymmetric field theory  $\mathcal{N}=4$ SYM in four space-time dimensions, and the type IIB string theory in $AdS_5\times S^5$ \cite{Maldacena:1997re,Aharony:1999ti}.  In top-down AdS/QCD models,  the energy scales are fixed by probe branes located in the bulk geometry. The presence of these probe branes in the AdS bulk breaks conformal symmetry and set the energy scales in the boundary theory \cite{Karch:2002sh,Sakai:2004cn,Sakai:2005yt}.  On the other hand, bottom-up AdS/QCD models implement deformations in the bulk geometry directly associated with observed phenomena in hadronic physics.  The most popular bottom-up AdS/QCD models are the hard wall  \cite{Polchinski:2001tt,BoschiFilho:2002ta,BoschiFilho:2002vd} and the soft wall \cite{Karch:2006pv}. The soft wall model is particularly interesting for investigating the radial excitations of mesons since it predicts a linear Regge trajectory for the hadron masses. Bottom-up AdS/QCD models have been systematically applied in the description of the spectrum of  mesons \cite{Karch:2006pv, Grigoryan:2007my,Erdmenger:2007cm,Colangelo:2008us, BallonBayona:2009ar, Cotrone:2010bv} and in particular for heavy  quarkonia \cite{Kim:2007rt,Grigoryan:2010pj,Li:2015zda,Braga:2015lck}. Heavy quark potentials have been analyzed for different botton-up AdS/QCD models, finding in all cases the linear behaviour for large separation \cite{BoschiFilho:2004ci,BoschiFilho:2006pe,Andreev:2006ct,Andreev:2006nw,Colangelo:2010pe,Bruni:2018dqm,Diles:2018wbe}. 
 
 The observed decay constants of quarkonia S-wave states increase the difficulty in obtaining an accurate description of their spectrum. The challenge comes from the fact that decay constants decrease in a monotonic and non-linear way with excitation level. The hard-wall model predicts decay constants increasing with excitation level, while the soft-wall model (quadratic dilaton) predicts completely degenerate decay constants. This poor description of decay constants at zero temperature leads to bad results at finite temperature, such as the disappearance of the spectral peaks of the fundamental state at low temperatures \cite{Fujita:2009wc,Fujita:2009ca,Mamani:2013ssa}. A good description of decay constants in the vacuum is needed to get a consistent spectral analysis at finite temperature. Decay constant defines the strength of the resonances fixing the zero-temperature limit of the spectral function. 

 In Ref. \cite{Grigoryan:2010pj} it is proposed an holographic description of $c\bar{c}$ considering modifications in the holographic potential. These modifications lead to an improvement in the description of masses and decay constants of $J/\Psi,\Psi'$. However, the holographic potential of \cite{Grigoryan:2010pj}  does not capture the decrease in decay constants. An alternative proposal  is  to set up an ultraviolet scale by calculating correlation functions in an AdS slice at finite $z_{uv}$  \cite{Evans:2006ea,Afonin:2011ff,Afonin:2012xq,Braga:2015jca}. This ultraviolet cut-off results in decay constants that decrease with excitation level. However, this model predicts a small decrease in the excitation level than experimental data that shows a fast decrease. So, it captures the decrease in decay constants but not the correct slope. The problem of the slope in decay constants was circumvented in a different holographic model proposed in Ref. \cite{Braga:2017bml} and refined in Ref. \cite{Braga:2018zlu}. The holographic model of Ref. \cite{Braga:2018zlu} captures the correct observed spectrum of decay constants of either charmonium and bottomonium with good precision. This success in describing the decay constants does not extend to the mass spectrum.  An accurate description of the radial excitations of heavy quarkonia involves either the masses and the decay constants. Here we propose a holographic model that simultaneously describes the masses and decay constants of the radial excitations of charmonium and bottomonium. The predictions of our model agree with experimental data within an RMS error near to $6\%$ for charmonium and $7,2\%$ for bottomonium, providing a precise description of quarkonia spectroscopy at zero temperature. We consider the effects of hot plasma on quarkonia states and use our model to compute in-medium spectral functions. We propose a prescription for background subtraction, isolating the contribution of the quasi-particle states in the spectral function from the medium effects. The melting temperatures of $J/\Psi,\Psi',\Upsilon,\Upsilon'$ are estimated and their thermal masses analyzed. 
 
The paper is organized as follows: in Section \ref{Holograph}, we motivate and present the dilaton that defines our holographic model. In Section \ref{zero-T}, we describe precisely the spectrum of masses and decay constants of charmonium and bottomonium. In Section \ref{finite-T} we consider our model at finite temperature: we discuss the confinement/ deconfinement phase transition, compute finite temperature spectral functions of $c\bar{c}$ and $b\bar{b}$ and analyse the quasi-particle states associated with the resonance peaks. In section \ref{Breit-Wigner} we perform the Breit-Wigner analysis to the holographic spectral densities calculated for heavy quarkonia. Finally, we elaborate in Section \ref{con} the main conclusions of this work.

\section{Holographic Model}\label{Holograph}
In the context of the AdS/QCD bottom-up models, heavy vector quarkonium is described as an abelian massless bulk gauge field.  This affirmation follows from the standard field/operator duality \cite{Aharony:1999ti}. Recall the scaling dimension of the source operators creating mesons at the conformal boundary defines the dual bulk field mass, according to the relation:

\begin{equation}
    M_5^2\,R^2=(\Delta-S)(\Delta+S-4),
\end{equation}

\noindent where $S$ is the meson spin, and $R$ is the AdS radius.  This relation defines a primitive notion of \emph{hadronic identity} since their corresponding bulk mass will categorize the dual hadronic states defined by the boundary source operator. In the case of \emph{any}  $q\,\bar{q}$ vector meson state, it is generated by structures with $\Delta=3$, implying $M_5^2\,R^2=0$. Thus, the action in the bulk space is given by

\begin{equation}
I_\text{Vector $Q\bar{Q}$}=-\frac{1}{4\,g_5^2}\,\int{d^5x\,\sqrt{-g}\,e^{-\Phi(z)}}\,g^{mp}\,g^{nr}\,F_{mn}\,F_{pr},    
\end{equation}

\noindent where $g_5$ is a constant that fixes units in the action given above and $F_{mn}$ is the field strength. This coupling is calculated from the large $q^2$ behavior of the holographic vector two-point functions \cite{Erlich:2005qh}.  The geometrical background is either AdS$_5$ or AdS$_5$ BH, depending on whether we are at zero or finite temperature. We will postpone this discussion  to the next section. 
Independent of the geometry, the equations of motion for the bulk gauge fields are

\begin{equation}
\frac{1}{\sqrt{-g}}\,\partial_n\left[\sqrt{-g}\,e^{-\Phi(z)}\,g^{np}\,g^{mr}\,F_{pr}\right]=0.    
\end{equation}

Confinement  in this model is induced \emph{via} the static dilaton field $\Phi(z)$. In the standard AdS/QCD softwall model, characterized by the static quadratic dilaton, large $z$ behavior guaranteed the emergence of linear radial Regge trajectories. However, it does not properly describe the meson decay constants since they are expected to decrease with the radial excitation number $n$. The softwall model calculation brings degenerate decay constants for $n$. 

A lesson learned from  \cite{MartinContreras:2019kah} was that decay constants depend on the low $z$ limit behavior of the AdS/QCD model at hand. We can modify this behavior by two possible forms: by deforming the background \cite{Braga:2015jca,Braga:2015lck} or by introducing terms in the dilaton that becomes relevant at low $z$ \cite{Braga:2017bml, Braga:2018zlu}. The resulting Regge trajectories are still linear, and the decay constant behavior is corrected. 

On the experimental side, these sorts of linear Regge trajectories describe better the light sector. Nevertheless, in the heavy one, the linear approximation does not seem to fit the available experimental data. By looking closely at the heavy quarkonium radial trajectories, we observed linearity in the highly excited states. On  the other side, the linear spectrum deviate due to the heavy constituent quark mass effect in the meson. This picture can be seen from the angular quantization of the string \cite{Afonin:2014nya} or the Bethe-Salpeter analysis \cite{Chen:2018nnr} by writing the radial trajectory as

\begin{equation}
(M_n-m_{Q_1}-m_{Q_2})^2=a(n+b),    
\end{equation}

\noindent where $a$ is a universal slope and $b$ is related to the mesonic quantum numbers. Therefore, nonlinearities emerge when the constituent quark mass comes to play. The nonlinear trajectories can be written in general as 

\begin{equation}
M^2=a(n+b)^\nu.     
\end{equation}

In a recent work \cite{MartinContreras:2020cyg}, these nonlinear Regge trajectories were described in the context of bottom-up holographic QCD. The main idea behind this model is that the inclusion of quark constituent masses implies deviation from the quadratic behavior imposed on the static dilaton.  This model successfully described vector mesons in the light unflavored, strange, heavy-light, and heavy sectors. 

This nonquadratic and the softwall model dilaton share the same low $z$ behavior. Therefore, in the nonquadratic context, the decay constants do not behave following the phenomenological constraints. An attempt to circumvent this drawback is by adding the proper low $z$ behavior that captures the expected decay constants phenomenology. Therefore we propose the following nonquadratic dilaton 

\begin{equation}\label{hybrid-dilaton}
\Phi(z)=\left(\kappa\,z\right)^{2-\alpha}+ M\,z\,+\text{tanh}\left[\frac{1}{M\,z}-\frac{\kappa}{\sqrt{\Gamma}}\right],    
\end{equation}

\noindent where the low $z$ contributions written above were read from \cite{Braga:2018zlu}.  The parameters $\kappa$, $M$ and $\sqrt{\Gamma}$ are energy scales controlling the slope and the intercept, whereas $\alpha$ is dimensionless and measures the constituent quark mass effect in the heavy meson, as it was introduced in  \cite{MartinContreras:2020cyg}.  

In the later sections, we will discuss the application of this dilaton for charmonium and bottomonium systems at zero and finite temperature.

\section{zero temperature}\label{zero-T}
In the case of zero temperature, the geometrical background is given by the Poincaré patch

\begin{equation}
dS^2=g_{mn}\,dx^m\,dx^n=\frac{R^2}{z^2}\left[dz^2+\eta_{\mu\,\nu}\,dx^\mu\,dx^\nu\right],    
\end{equation}

\noindent with the signature $\eta_{\mu\nu}=\text{diag}(-1,1,1,1)$ and $z\in (0,\infty)$. 

Following the AdS/CFT methodology, we will write the Fourier transformed bulk vector field in terms of the bulk modes $\psi(z,q)$ and the boundary sources  as

\begin{equation}
A_\mu(z,q)=A_\mu(q)\,\psi(z,q),    
\end{equation}

\noindent where we have implicitly imposed the gauge fixing $A_z=0$. We use the $z$ component of the equations of motion, $\partial_z(\partial_\mu A^\mu)=0$, and the  Lorentz gauge in the boundary to set $\partial_\mu A^\mu=0$ everywhere. These definitions yield the following equations for the eigenmodes

\begin{equation}\label{eqn-HQ}
\partial_z\left[e^{-B(z)}\,\partial_z\,\psi_n(z,q)\right]+(-q^2)\,e^{-B(z)}\,\psi_n(z,q)=0.    
\end{equation}

\noindent where we have defined  the background information $B(z)$ function as

\begin{equation}
B(z)=\Phi(z)-\text{log}\left(\frac{R}{z}\right).    
\end{equation}

Confinement emerges in this model by the effect of the dilaton field that induces a holographic confining potential. We apply the Bogoliubov transformation $\psi(z)=e^{B(z)/2}\,\phi(z)$ to the expression \eqref{eqn-HQ} obtaining a Schrodinger-like equation defined as

\begin{equation}
-\phi_n''(z)+U(z)\,\phi_n(z)=M_n^2\,\phi_n(z),    
\end{equation}

\noindent where $M_n^2=-q^2$ defines the meson spectrum, and the holographic potential is constructed in terms of the derivatives of the $\Phi(z)$ dilaton field in eqn. \eqref{hybrid-dilaton} as follows 

\begin{equation}\label{holo-pot-zero}
U(z)= \frac{3}{4\,z^2}+\frac{\Phi'(z)}{2\,z}+\frac{1}{4}\Phi'(z)^2-\frac{1}{2}\Phi''(z).
\end{equation}

By solving the Schrodinger-like equation numerically, we obtain the associated bulk modes and the holographic mass spectrum. The results for charmonium and bottomonium, with the corresponding parameter fixing,  are summarized in tables \ref{tab:one}
and \ref{tab:two}. 

\begin{center}
\begin{table*}[t]
    %\centering
    \begin{tabular}{||c||c||c|c|c||c|c|c||}
    \hline
    \multicolumn{8}{||c||}{\textbf{Charmonium States $I^G(J^{PC})=0^+(1^{--})$}}\\
    \hline
    \hline 
\multicolumn{3}{||c|}{\textbf{Parameters}:}&\multicolumn{5}{c||}{$\kappa= 1.8$ GeV, $M=1.7$ GeV, $\sqrt{\Gamma}=0.53$ GeV and $\alpha=0.54$} \\   
\hline
\hline
\textbf{$n$} & \textbf{State} & \textbf{$M_\text{Exp}$ (MeV)} & \textbf{$M_\text{Th}$ (MeV)} &\textbf{\%$M$}&\textbf{$f_\text{Exp}$ (MeV)} &\textbf{$f_\text{Th}$ (MeV)}&\textbf{\%$f$}\\
\hline
\hline
$1$ & $J/\psi$ & $3096.916\pm0.011$ & $3140.1$ & $1.42$ & $416.16\pm 5.25$ & $412.4$ &$1.4$\\
$2$ & $\psi(2S)$ & $3686.109\pm 0.012$ & $3656.9$& $0.9$ & $296.08\pm 2.51$ & $272.7$ & $8.0$\\
$3$ & $\psi(4040)$ & $4039\pm 1$& $4055.7$ & $0.4$ & $187.13\pm7.61$& $201.8$ &$7.8$\\
$4$ & $\psi(4415)$ & $4421\pm4$ & $4376$& $0.9$ & $160.78\pm9.70$ & $164.1$ & $2.0$\\
\hline 
\hline 
\multicolumn{4}{||c|}{\textbf{Nonlinear Regge Trajectory:}}&\multicolumn{4}{|c||}{$M_n^2=8.097(0.39+n)^{0.58}$GeV$^2$ with $R^2=0.999$}\\
\hline
\hline
    \end{tabular}
    \caption{Summary of results for charmonium states. $M_{Th}$ and $f_{Th}$ are calculated with the parameters mentioned on header, and corresponding errors appear in columns $\% M$ and $\% f$. Experimental results are read from PDG \cite{Tanabashi:2018oca} and total error is $\delta_\text{RMS}=6.0\,\%$.  The Regge trajectories are also presented.}
     \label{tab:one}
\end{table*}    
\end{center} 

\begin{center}
\begin{table*}[t]
    %\centering
    \begin{tabular}{||c||c||c|c|c||c|c|c||}
    \hline
    \multicolumn{8}{||c||}{\textbf{Bottomonium States $I^G(J^{PC})=0^+(1^{--})$}}\\
    \hline
    \hline 
\multicolumn{3}{||c|}{\textbf{Parameters}:}&\multicolumn{5}{c||}{$\kappa= 9.9$ GeV, $M=2.74$ GeV, $\sqrt{\Gamma}=1.92$ GeV and $\alpha=0.863$} \\   
\hline
\hline
\textbf{$n$} & \textbf{State} & \textbf{$M_\text{Exp}$ (MeV)} & \textbf{$M_\text{Th}$ (MeV)} &\textbf{\%$M$}&\textbf{$f_\text{Exp}$ (MeV)} &\textbf{$f_\text{Th}$ (MeV)}&\textbf{\%$f$}\\
\hline
\hline
$1$ & $\Upsilon(1S)$ & $9460.3\pm0.26$ & $9506.5$ & $0.5$ & $714.99\pm2.40$ & $718.8$ &$0.5$\\
$2$ & $\Upsilon(2S)$ & $10023.26\pm0.32$ & $9892.9$& $1.0$ & $497.37\pm2.23$ & $575.7$ & $16$\\
$3$ & $\Upsilon(3S)$ & $10355.2\pm0.5$ & $10227.2$ & $1.2$ &$430.11\pm1.94$ & $413.0$ &$4.0$\\
$4$ & $\Upsilon(4S)$ & $10579.4\pm1.2$ & $10497.5$& $0.8$ & $340.65\pm9.08$ & $324.3$ & $4.8$\\
$5$ & $\Upsilon(10860)$ & $10889.9^{+3.2}_{-2.6}$& $10721.5$ & $1.5$ & -- &--&--\\
$6$ & $\Upsilon(11020)$ & $10992.9^{+10.0}_{-3.1}$ & $10912.7$ & $0.7$ & --&--&--\\
\hline 
\hline 
\multicolumn{4}{||c|}{\textbf{Nonlinear Regge Trajectory:}}&\multicolumn{4}{|c||}{$M_n^2=7.376(1.31+n)^{0.24}$GeV$^2$ with $R^2=0.999$}\\
\hline
\hline
    \end{tabular}
    \caption{Summary of results for bottomonium states. $M_{Th}$ and $f_{Th}$ are calculated with the parameters mentioned on header, and corresponding errors appear in columns $\% M$ and $\% f$. Experimental results are read from PDG \cite{Tanabashi:2018oca} and total error is $\delta_\text{RMS}=7.2\,\%$.  The Regge trajectories are also presented.}
     \label{tab:two}
\end{table*}    
\end{center} 

In the case of electromagnetic decay constants $f_n$, they arise as the residues of the expansion in poles $-q^2=M_n^2$ of the two-point function, defined from the correlator of two electromagnetic currents:

\begin{eqnarray}\notag
\Pi_{\mu\nu}(q^2)&=&i\,\int{d^4x\,e^{i\,q\cdot x}\langle 0\left|\mathcal{T}\left\{j_\mu(x)\,j_\nu(0)\right\}\right|0\rangle}\\ \notag
&=&\left(q_\mu\,q_\nu-q^2\,\eta_{\mu\nu}\right)\,\Pi(-q^2)\\
&=&\left(q_\mu\,q_\nu-q^2\,\eta_{\mu\nu}\right)\,\sum_{n}{\frac{f_n^2}{-q^2-M_n^2+i\,\varepsilon}}.
\end{eqnarray}

\noindent The tensor structure written in parentheses is nothing else than the transverse projector, coming from the fulfillment of the Ward-Takahashi identities. 

The importance of the two-point function relies on the description of the intermediate hadronic states that appear in scattering processes involving hadrons. Decay constants measure the probability of finding such states in the collision final products.

In the case of heavy quarks, the  electromagnetic quark currents $e\,\bar{c}\,\gamma_\mu\,c$ and $e\,\bar{b}\,\gamma_\mu\,b$ creates the $J/\psi$ and $\Upsilon$ mesons respectively.  At the physical level, these mesonic vector states appear as observed resonances in the $e^+\,e^-$ annihilation process when the center of mass energy is around the mass of the corresponding mesonic states. Therefore, these states are expected to be also poles in the two-point function.

Experimentally, decay constants are measured from the vector meson decaying process $V\to e^+\,e^-$, according to the expression:

 \begin{equation}
    f_n^2=\frac{3\,M_n\,\Gamma_{V\to e^+e^-}}{4\,\pi\,\alpha^2_\text{em}\,C_V^2},    
\end{equation}

\noindent where $\Gamma_{V\to e^+\,e^-}$ is the heavy vector decay width, and $C_V$ stands for the heavy quark electromagnetic charge contribution to the meson, i.e., $C_{J/\psi}=2/3$ and $C_{\Upsilon}=1/3$. 

The holographic dual of the two-point function is determined from the on-shell boundary action \cite{Karch:2006pv}. Following the field/operator duality, the holographic two-point is written as

\begin{equation}\label{2-point-f}
\Pi\left(-q^2\right)=-\left.\frac{e^{-B\left(z\right)}}{g_5^2\,(-q^2)}\,\partial_z\,V\left(z,q\right)\right|_{z\to0},    
\end{equation}
where $V(z,q)$ is the bulk-to-boundary propagator.  It is straightforward to prove that this object can be written in terms of the normalizable modes $\psi(z,q)$ by using the Green's function associated with the equations of motion \eqref{eqn-HQ}.  In work \cite{MartinContreras:2019kah}, authors follow this path deriving a general expression for the decay constants calculated for any general AdS/QCD model depending only on the value of the quotient $\psi(z,q)/z^2$ and the dilaton at the conformal boundary

\begin{equation}\label{decay-cons-2}
    f_n^2=\frac{1}{M_n^2\,g_5^2}\,\lim_{z\to0}{\,e^{-2\,\Phi(z)}\,\left|\frac{2\,\psi_n(z,q)}{z^2}\right|^2}.
\end{equation}

Let us stop here and see how the decay constants are calculated in the soft wall model, i.e., static and quadratic dilaton. Following \cite{Karch:2006pv}, we see that the mass spectrum has the linear structure $M_n^2=4\,k^2(n+1)$, with $k$ being the dilaton slope.  The eigenfunctions are defined in terms of Laguerre associated polynomials

\begin{equation}
\psi_n(z)=\sqrt{\frac{2\,k^4\,n!}{(n+1)!}}\,z^2\,L_n^1(k^2\,z^2),   
\end{equation}

\noindent therefore, the decay constants follow from eqn. \eqref{decay-cons-2} yielding

\begin{equation}
f_n^2=\frac{F_n^2}{M_n^2}=\frac{1}{4\,g_5^2\,k^2\,(n+1)}\times\,\frac{8\,k^4\,(n+1)!}{n!}=
\frac{2\,k^2}{g_5^2},
\end{equation}

\noindent where we have used the asymptotic form of the Laguerre associated polynomials when $z\to 0$. Therefore, we can conclude that decay constants are degenerate in the softwall model. If we do similar computations in the hardware model context \cite{BoschiFilho:2002vd}, they will lead to increasing decays $f_n$ with the excitation number $n$. This drawback can be avoided by deforming the low $z$ limit in the static dilaton, as it was first noticed by Braga et al. \cite{Braga:2016oem}. We will extend this idea in the context of non-quadratic dilatons.

The numerical results for the charmonium and bottomonium decay constants, calculated in the deformed non-quadratic dilaton context, are summarized in tables \ref{tab:one}
and \ref{tab:two}. The deviations presented in the caption of tables \ref{tab:one} and \ref{tab:two} represent the difference between the theoretical prediction and the most probable value of a given experimental measure. The total deviation $\delta_{RMS}$ is defined as

\begin{equation}
\delta_\text{RMS}=\sqrt{\frac{1}{N-N_p}\sum_i^N\left(\frac{\delta\,O_i}{O_i}\right)^2},  \label{delta}
\end{equation}

\noindent where $O_i$ is a given experimental measure with $\delta\, O_i$ defining the deviation of the theoretical value from the experimental one, $N_p$ is the number of model parameters, and $N$ the total number of available observables.\par

\section{Finite temperature}\label{finite-T}
For the finite-temperature extension, we will consider the heavy quarkonium system living in a thermal bath, addressed by a colored plasma. Holographically, we will deal with vector bulk field living in an AdS-Schwarzschild black hole background, described by the metric

\begin{equation}\label{AdS-BH-Schw}
dS_\text{AdS-Schw}^2=\frac{R^2}{z^2}\left[\frac{dz^2}{f(z)}-f(z)\,dt^2+d\vec{x}\cdot d\vec{x}\right],
\end{equation}

with the blackening factor defined as 

\begin{equation}
f(z)=1-\frac{z^4}{z_h^4}.    
\end{equation}

\noindent where $z_h$ is the event horizon locus. 

The description of heavy quarkonium at finite temperature in the context of the softwall model was developed in \cite{Fujita:2009ca}. However, as it was discussed in \cite{Vega:2016gip,Vega:2017dbt,Vega:2018dgk}, by analyzing the holographic potential in the context of Bogoliubov transformations and tortoise coordinates, the mesonic melting temperature appears to be too low as the ones expected from lattice QCD. This bad behavior is attached to the holographic decay constant description in the softwall model, where these objects are degenerate. This affirmation is sustained by the thermal analysis of the hadronic part of  the two-point function \cite{Dominguez:2009mk,Dominguez:2013fca}. For instance,  the hadronic spectral density calculated from thermal sum rules

\begin{equation}
\left.\frac{1}{\pi}\mathbb{I}\text{m}\,\Pi(s,T)\right|_\text{hadron}=\frac{f_n^2\,M_n(T)^3\,\Gamma_n(T)}{[s-M_n^2(T)]^2+M_n^2(T)\,\Gamma_n(T)^2},    
\end{equation}

\noindent establishes the formal dependence of the melting process with the decay constant. 

This softwall model issue was circumvented by introducing low $z$ modifications into the model, as it was done in \cite{Braga:2016wkm}. Therefore, it is natural to suppose that this hybrid dilaton should exhibit the expected raising in the melting temperatures in agreement with phenomenology.  

Let us focus on reviewing the holographic description of the heavy quarkonium.  Our starting point is the calculation of the hadronic spectral density. To do so, we will follow the Minkowskian prescription given by \cite{Son:2002sd}. Let us perform the variable change $z=z_h\,u$ in the metric \eqref{AdS-BH-Schw} in order to fix the horizon locus at $u=1$.  We will also fix $-q^2=\omega^2$ in our analysis. 

\subsection{Confinement/Deconfinement phase transition}

In the boundary gauge theory, the formation of a deconfined plasma is holographically described via the Hawking-Page phase transition in the dual geometry \cite{Hawking:1982dh, Herzog:2006ra}. On the gauge theory side, above the critical temperature, $T_c$, the fundamental quarks and gluons inside the colorless matter are allowed to walk away from its partners, forming a plasma of deconfined colored particles. It is usually considered that the light vector meson dominates the deconfinement transitions. That is, the medium is formed when the light quarks can escape from the hadrons. Consequently, we use the light meson spectrum to fix the energy scales governing the confinement/deconfinement transition.  

The observed spectrum of radial excitations of the $\rho$ meson includes the masses of the first five radial excitations, and the decay constant of the ground state \cite{Tanabashi:2018oca}. It is important to mention that additional scales in the model encode heavy quarkonia properties and bring no particular advantages in describing the light meson spectrum. In particular, for light mesons, the parameter $\alpha$ in eq.(\ref{hybrid-dilaton}) is set to vanish. The observed spectrum of the radial excitations of the $\rho$ meson are reasonable fitted using the model parameters $\kappa=0.6$ GeV, $M=0.06$ GeV, $\sqrt{\Gamma}=0.02$ GeV. Using these parameters to fix the dilaton profile, we compute the gravitational on-shell action of the AdS-Schwarzschild black hole geometry and the thermal AdS geometry. The normalized difference is then obtained as
 
 \begin{equation}
\Delta S = \int_\epsilon^{z_h}dz\frac{e^{-\Phi(z)}}{z^5} - \sqrt{f(\epsilon)}\int_\epsilon^\infty dz\frac{e^{-\Phi(z)}}{z^5}.     
 \end{equation}
 
\begin{figure*}
    \includegraphics[width=3.4 in]{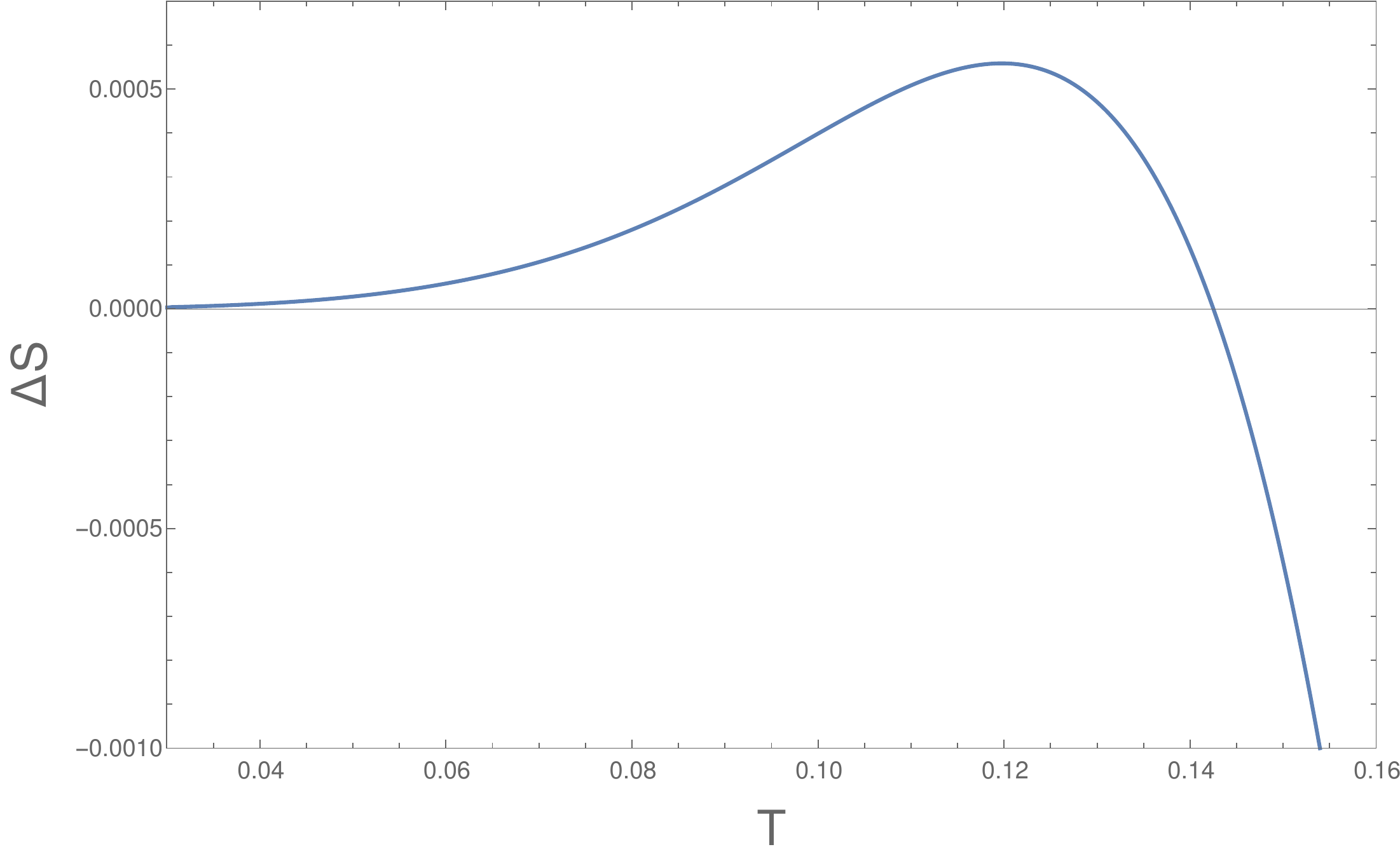}
\caption{The difference between the on-shell gravitational action of AdS-Schwarzschild and Thermal AdS geometries is depicted as a function of temperature in GeV. The intersection with the horizontal axis gives the critical temperature of the deconfinement transition.}
\label{Tc}
\end{figure*}

We show in Figure \ref{Tc} the difference in action as a function of temperature. In the region where $\Delta S$ is positive, the thermal AdS is stable. In the region with $\Delta S$ is negative, the black hole is stable. The condition $\Delta S=0$ defines the critical temperature, and we obtain 

   \begin{equation}
       T_c=142\,~ \textrm{MeV.}
   \end{equation}

 There are two important comments to make at this point. First, using the $\rho$ meson spectrum to fix model parameters is a particular choice. As it was recently pointed out in \cite{Afonin:2020gka}, the definition of $T_c$ through a Hawking-Page transition is model depending. The same authors performed a similar calculation considering the gluon condensate obtaining a critical temperature of $156$ MeV \cite{Afonin:2020crk}. Second, the phase transition associated with QGP formation in heavy-ion collisions is more likely a continuous crossing over than an abrupt transition \cite{Aoki:2006we}. However, the present computation of $T_c$ has no intention of dealing with these subtleties. The critical temperature we obtain ($T_c=142$ MeV) is consistent with the present holographic model and will be adopted from now on.

\subsection{Spectral density}
The holographic spectral density comes from the thermal Green's function. We define the bulk-to-boundary propagator in momentum space $V_\mu(z,q)=V(z,q)V^0_\mu(q)$, where $V^0_\mu(q)$ is the source at the boundary.
According to the Minkowskian prescription, this correlator is written in terms  of the derivatives of the bulk-to-boundary propagator $V(z,q)$ as  

\begin{equation}\label{Green-F-Ther}
G_R(\omega)=-\left.\frac{2}{z_h\,\mathcal{N}}\,\,e^{-B(u)}\,f(u)\,V(u,-\omega)\,\partial_u\,V(u,\omega)\right|_{u=0}.
\end{equation}

The spectral density, according to the Kubo relations, is written as the imaginary part of the Green's function

\begin{equation}\label{spectral.density}
\rho(\omega)=-\mathbb{I}\text{m}\,G_R(\omega).  
\end{equation}

The bulk-to-boundary propagator obeys the bulk spatial vector equation of motion

\begin{equation}\label{eom-bvp}
\partial_u\left[e^{-B(u)}\,f(u)\,\partial_u\,V(u,\omega)\right]+\frac{z_h^2\,\omega^2}{f(u)}e^{-B(u)}\,V(u,\omega)=0.    
\end{equation}

Although we are at finite temperature, the bulk-to-boundary propagator still preserves its properties at the conformal boundary. If this is not guaranteed, the field/operator duality does not hold anymore. Recall that at the conformal boundary, we require that $V(u\to0)\to1$. On the other side, we also need that $V(u,\omega)$ obeys the out-going boundary condition $\phi_-(u)$, defined as

\begin{equation}
\phi_-(u)=\left(1-u\right)^{-\,i\frac{\omega\,z_h}{4}}    
\end{equation}

These conditions define the procedure to compute the spectral density. We will follow the method depicted in \cite{Teaney:2006nc, Fujita:2009ca,Miranda:2009uw,Fujita:2009wc}.  Our starting point is writing a general solution $v(u)$ for the Eqn. \eqref{eom-bvp} in terms of the normalizable $\psi_0(u)$ and non-normalizable $\psi_1(u)$, that form a basis, in the following form

\begin{equation}
v(u)=A\,\left[\psi_1(u)+\frac{B}{A}\,\psi_0(u)\right],    
\end{equation}

\noindent such that the bulk-to-boundary propagator is written as $V(\omega,u)=A^{-1}\,v(u)$, and satisfying the asymptotic solutions near the conformal boundary

\begin{eqnarray}
\psi_0(u\,\omega)&=&\frac{2}{\omega\,z_h}\,u\,J_1(\omega\,z_h\,u)\\
\psi_1(u\,\omega)&=&-\frac{\pi\,\omega\,z_h}{2}\,u\,Y_1(\omega\,z_h\,u)
\end{eqnarray}

After replacing this solution into the Green's function definition we obtain

\begin{equation}
G_R(\omega)=-\left.\frac{2\,R}{z_h\,\mathcal{N}}\left[\frac{B}{A} -\frac{\omega^2\,z_h^2}{2}\,\text{log}\,\left(\frac{e^{\gamma_e}\,\varepsilon\,\omega\,z_h}{2}\right)\,\varepsilon^2\right] \right|_{\varepsilon\to0}
\end{equation}

Finally, the spectral density is written as the imaginary part of the Green's function

\begin{eqnarray}\notag
\rho(\omega)&=&-\mathbb{I}\text{m}\,G_R(\omega)\\ \label{spectral-density-1}
&=&\frac{2\,R}{z_h\,\mathcal{N}}\,\mathbb{I}\text{m}\frac{B}{A}.
\end{eqnarray}

\begin{center}
\begin{figure*}
  \begin{tabular}{c c}
    \includegraphics[width=3.4 in]{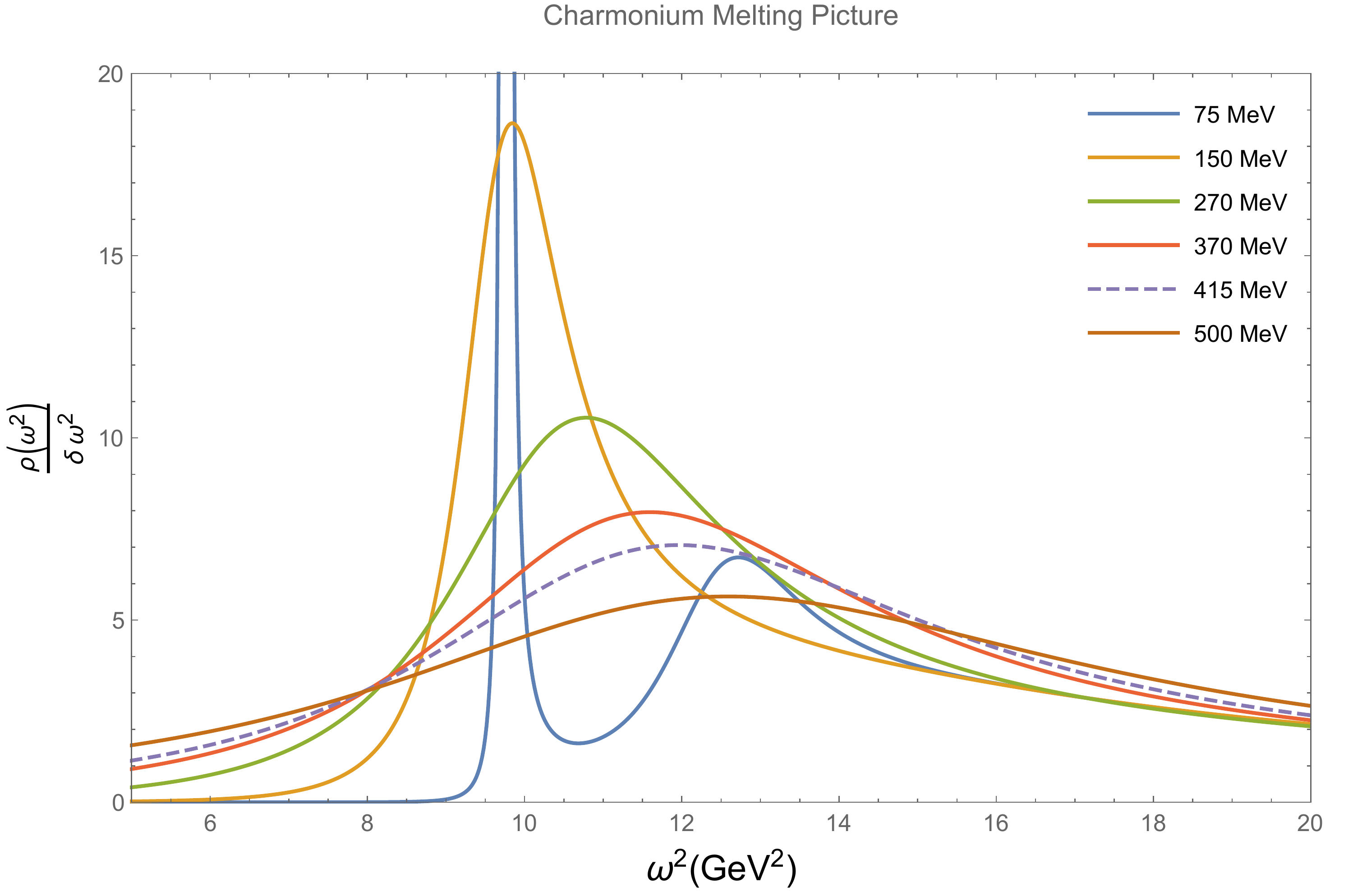}
    \includegraphics[width=3.4 in]{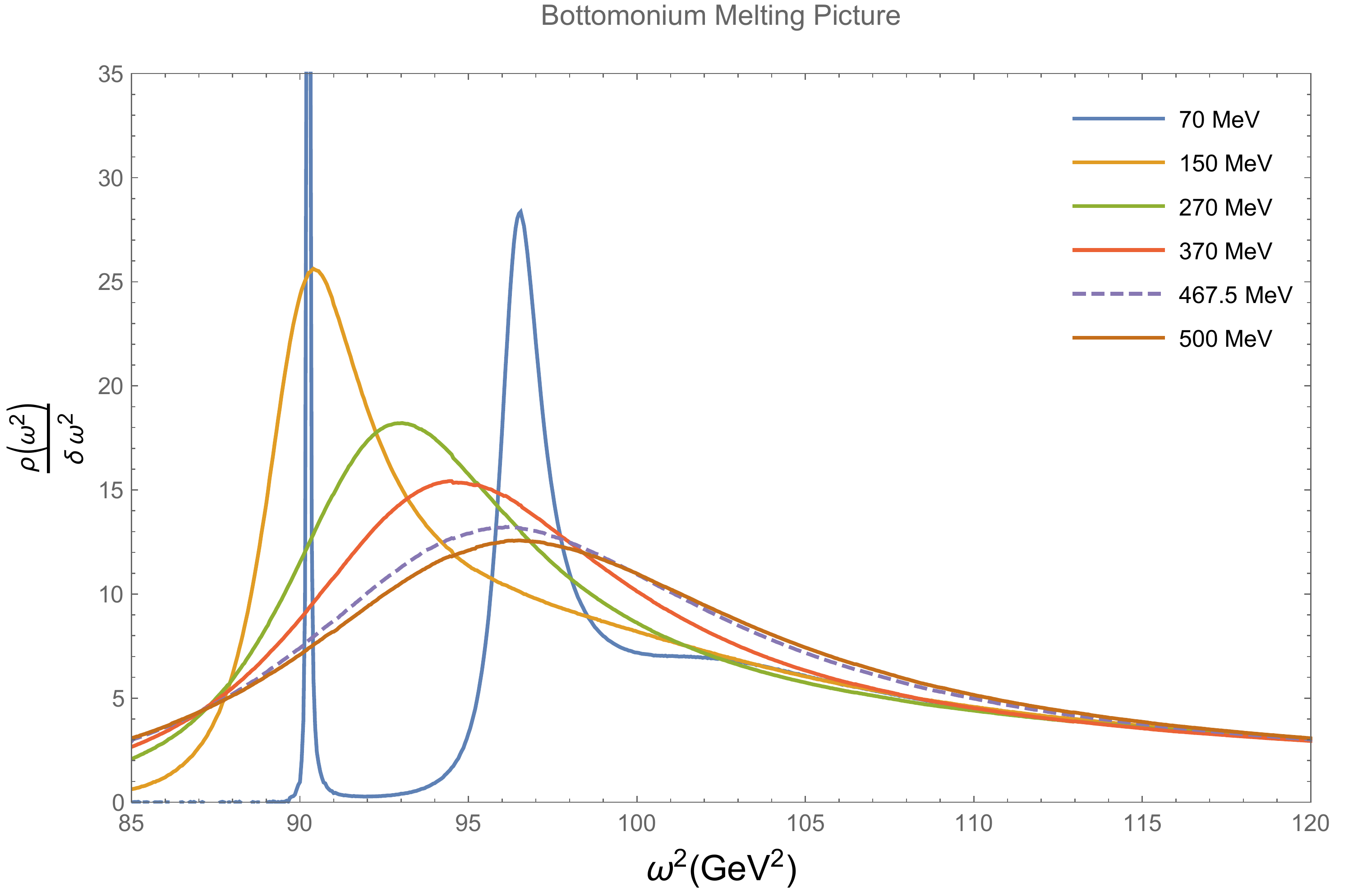}
  \end{tabular}

\caption{This figure describes the spectral density for charmonium (left panel) and bottomonium (right panel) calculated using Eqn. \eqref{spectral-density-1}, depicting the melting process.Dashed lines corresponds to the melting temperature in each case.}
\label{fig:one}
\end{figure*}
\end{center}

Numerical results for the spectral density calculated for charmonium and bottomonium system are shown in Fig. \ref{fig:one}. %Table \ref{tab:three} summarizes the melting temperature observed with the spectral density. 

\subsection{Thermal holographic potential}
Another essential quantity that carries valuable information about the heavy quarkonium thermal picture is the thermal potential. At zero temperature case, the potential translates the dilaton effect into the holographic confinement. Holographic mesonic states appear as eigenfunctions of this potential. 

The thermal dissociation of mesons is connected with the holographic potential. In  \cite{Vega:2018dgk}, this idea was discussed in the context of softwall-like dilatons that vanish at the conformal boundary. In this proposal, the melting is characterized by the disappearance of the potential well. At zero temperature, the dilaton vanishes near the boundary, and the potential holographic displays one single minimum that is global at zero temperature. The disappearance of the global minimum of the holographic potential encodes the information of meson dissociation.

In this work,  we consider a dilaton that does not vanish near the boundary. This dilaton field, given in Eqn. \eqref{hybrid-dilaton} interpolates between linear and the deformed quadratic behavior, which induces a nonlinear spectrum. This dilaton also changes the global structure of the potential by introducing  local minima near the UV at zero temperature. As argued in \cite{Grigoryan:2010pj, MartinContreras:2019kah}, this UV deformation is needed in order to describe the proper phenomenological behavior the decay constants of the excited quarkonia states.  

It is expected that, at finite temperature, the holographic potential also has information about the melting process. To make a formal approach to this phenomenology, we apply  the Liouville (tortoise)  transformation. It transforms the equations of motion into a Schr\"odinger-like equation in terms of a Liouville (tortoise) coordinate $r^*$. The potential exhibits a barrier that decreases with the temperature, mimicking how the confinement starts to cease when the temperature rises. Following \cite{Vega:2018dgk}, one expect that the barrier disappears when all of the quarkonia states melt down into the thermal medium. However, the appearance of a local minima near $z=0$ can sustain the state after the disappearance of the barrier.

The Liouville transformation appears in the core of the Liouville theory of second-order differential equations. Given a differential equation, we can associate it with a differential diagonalizable operator. As a consequence, this operator will acquire a spectrum of eigenvalues and eigenfunctions. In the holographic case at hand, the associated potential is defined \emph{via} the transformation

\begin{equation}\label{tortoise-coor}
r^*(u)=z_h\,\int_0^u{\frac{d\,\xi}{1-\xi^4}}=\frac{z_h}{2}\left(\text{tan}^{-1}\,u+\text{tanh}^{-1}\,u\right).    
\end{equation}

The equations of motion \eqref{eom-bvp} transform into the following Schrodinger-like equation 

\begin{equation}
-\frac{d^2\,\phi(r^*)}{d\,r^{*2}}+U(r^*)\,\phi(r^*)=\omega^2\,z_h^2\,\phi(r^*),     
\end{equation}

with the following definitions

\begin{multline}\label{tortoise-pot}
U(r*)=f(u)^2\left[\frac{3}{4\,u^2}+\frac{\Phi'(u)}{2\,u}+\frac{\Phi'(u)^2}{4}-\frac{\Phi''(u)}{2}\right.\\
\left.-\frac{f'(u)}{2\,u\,f(u)}-\frac{f'(u)\,\Phi'(u)}{2\,f(u)}\right]    
\end{multline}
\begin{eqnarray}
\phi(r^*)&=&\psi(u)\,e^{\frac{1}{2}\,B(u)}\\
u&=&u(r^*).
\end{eqnarray}

\noindent where $u=u(r^*)$ is obtained by inverting the Liouville coordinate defined in Eqn. \eqref{tortoise-coor}.

\begin{center}
\begin{figure*}
  \begin{tabular}{c c}
    \includegraphics[width=3.4 in]{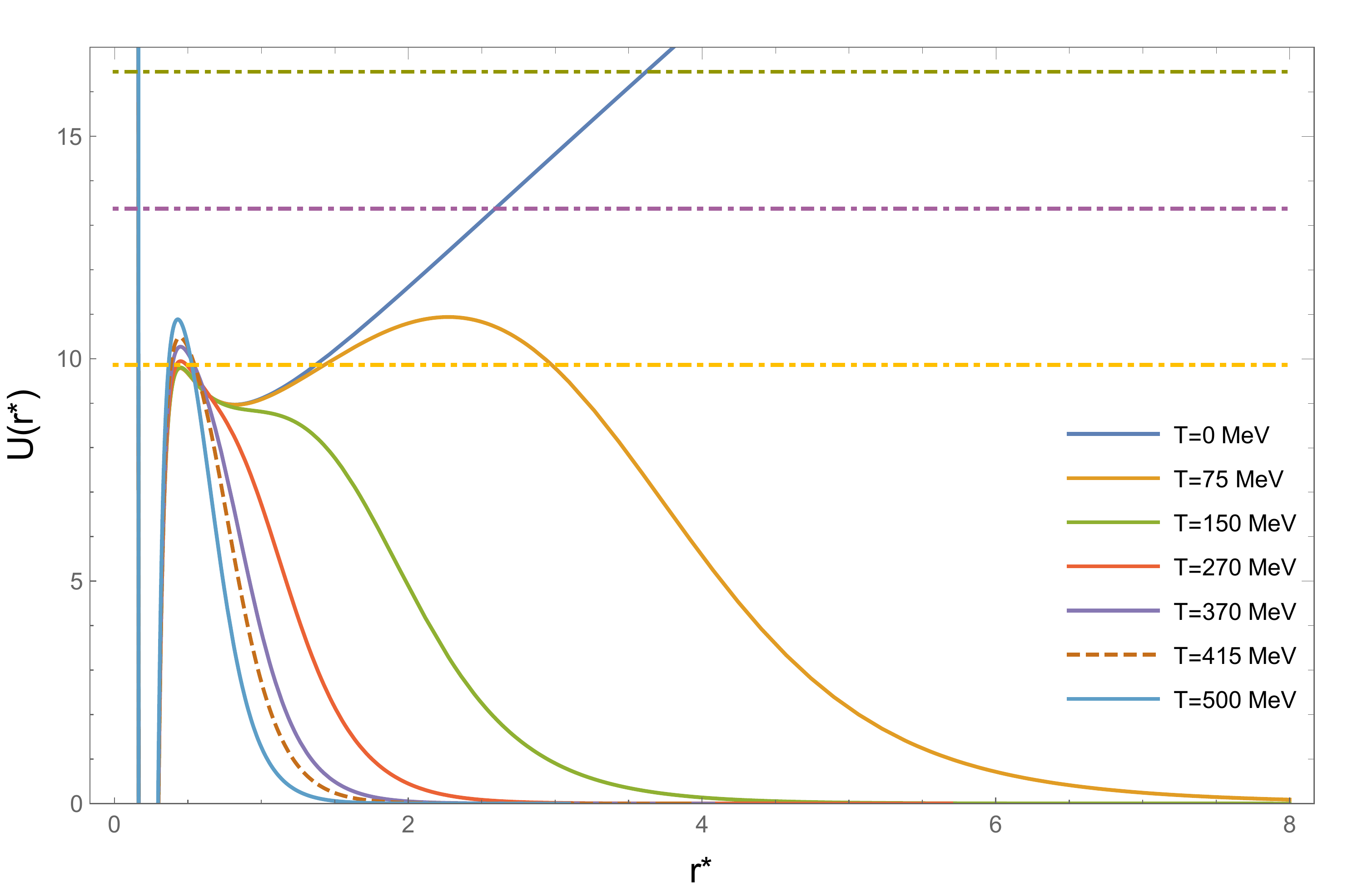}
    \includegraphics[width=3.4 in]{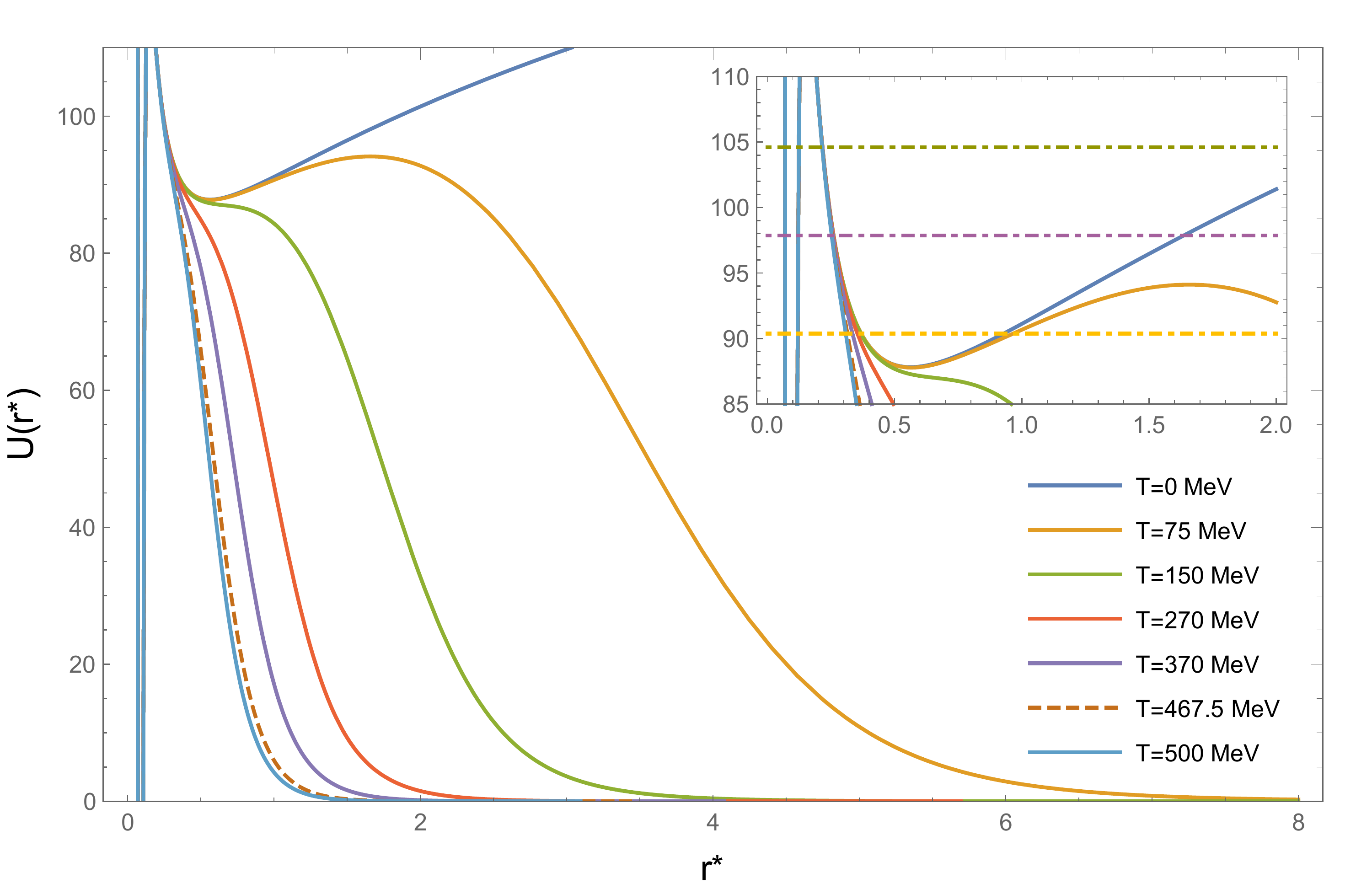}
  \end{tabular}

\caption{In this figure, we plot the holographic Liouville potential for charmonium (left panel) and bottomonium (right) panel. Also, we plot the first three masses calculated a zero temperature to illustrate the melting process. When the barrier decreases below the mass, we can consider that such a state had undergone a melting process. }
\label{fig:two}
\end{figure*}
\end{center}

In figure \ref{fig:two}, we depict the melting process from the Liouville potential for the heavy quarkonia.  In the zero temperature case, the potential reduces to the holographic one described in Eqn. \eqref{holo-pot-zero}.

The melting process in the present case is a two step process involving two different energy scales. The first step is the disappearance of the infra-red barrier when the temperature is increased above $T_c$ allowing for the bulk modes to be absolved by the event horizon. At this step all the excited states melts in the thermal medium. But this is not sufficient to state the melting of the ground state. The appearance of a deep, narrow and persistent well near $z=0$ produces a barrier greater them the mass of the ground state. The well is separated from the event horizon by a barrier which narrows with the raising of temperature. At the melting temperature the barrier is too narrow to hold the bulk wave packet, that escapes from the well and is absolved by the event horizon. A quantitative description of the tunneling process is not performed here and the melting temperature depicted in Figure (\ref{fig:two}) are obtained from the Breight-Wigner analysis performed in the next section.

\section{Breit-Wigner analysis}\label{Breit-Wigner}
Once the spectral functions are calculated, we will perform the Breit-Wigner analysis to discuss the thermal properties captured by the holographic model described above. This analysis allows extracting information about the meson melting process, as the temperature and the thermal mass shifting.  Recall that when a meson starts to melt, the resonance begins to broad (the width becomes large), and the peak height, which is proportional to the decay constant, decreases.  In other words, the mesonic couplings tend to zero as the temperature rises, implying these states ceased to be formed in the colored medium.

Therefore, comparing the peak height and the width size will be the natural form to define the meson melting temperature: the temperature at which the width size overcomes the peak high is where the meson starts to melt. This phenomenological landscape also comes in the context of pNRQCD at thermal equilibrium.

The next thing to consider is the background. These background effects observed in the spectral function come from continuum contribution, and they should be subtracted in order to isolate the Breit-Wigner behavior. The background subtraction methodology is not unique, and in general, is model depending. However, most of the authors define interpolation polynomials in terms o powers of $\omega^2$. See, for example, \cite{Colangelo:2009ra,Cui_2016} in the light scalar sector and \cite{Fujita:2009ca} for heavy vector quarkonium. In these references, authors worked with quadratic-like dilatons.

In our particular case, we will follow a different path:  we will consider the large $\omega^2$ behavior to deduce a background subtraction mechanism.  As ref. \cite{Grigoryan:2010pj} pointed it out,  in a conformal theory at short distances, we could expect that

\begin{equation}
    \lim_{\omega^2\to\infty}\frac{\rho(\omega^2)}{\omega^2}=\frac{\pi}{2\,g_5^2} \,\,\,\text{ i.e., a dimensionless constant},
\end{equation}

\noindent for the case of quadratic-like dilatons. The OPE-expansion of the 2-point function dictates this behavior, allowing the match between the bulk and the boundary theories. In the purely phenomenological sense, the existence of this dimensional constant is a signature of asymptotic freedom. Thus, the spectral function for these quadratic-like dilatons can be rescaled as 

\begin{equation}
\bar{\rho}(\omega^2)=\frac{\rho}{\omega^2},    
\end{equation}

\noindent in order to test the asymptotic freedom signature in the model. Therefore, if the rescaled spectral function behavior does not match this criterion, the model does not have a proper large $\omega^2$ limit compared with QCD. The softwall model with quadratic dilaton perfectly matches this condition.

Then, what happens when the model does not have a quadratic dilaton? To answer this question, we can go further by imposing the same asymptotic condition.  However, changing the quadratic structure on the dilaton will imply that the asymptotic behavior of the spectral function is different: it is still linear in $\omega^2$, but with a shifted value of the coupling $g_5$, defined at zero temperature from the holographic 2-point function. Thus, we suggest the following rescaling:

\begin{equation}
    \bar{\rho}(\omega^2)=\frac{\rho(\omega^2)}{\delta\,\omega^2},
\end{equation}

\noindent where $\delta$ is determined from the large $\omega^2$ behavior observed in the spectral function $\rho(\omega^2)$. From this rescaled spectral function, we will subtract the background effects and construct the Breit-Wigner analysis. For our practical purposes, we will write the Breit-Wigner distribution as 

\begin{equation}\label{BW-fit}
\bar{\rho}(\omega^2)=\frac{1}{2}\frac{A_0\,\omega^2_0\,\Gamma_0\,\omega^{a_0}}{(\omega^2-\omega^2_0)^2+\frac{\omega^2_0\,\Gamma^2_0}{4}},    
\end{equation}

\noindent where $A_0$, $a_0$ are fitting parameters, $\omega_0$ is the mesonic peak and $\Gamma_0$ is the decay width, proportional to the inverse of the meson life-time.

%%%% Here
\subsection{Background substraction}
\begin{center}
\begin{figure*}
  \begin{tabular}{c c}
    \includegraphics[width=3.4 in]{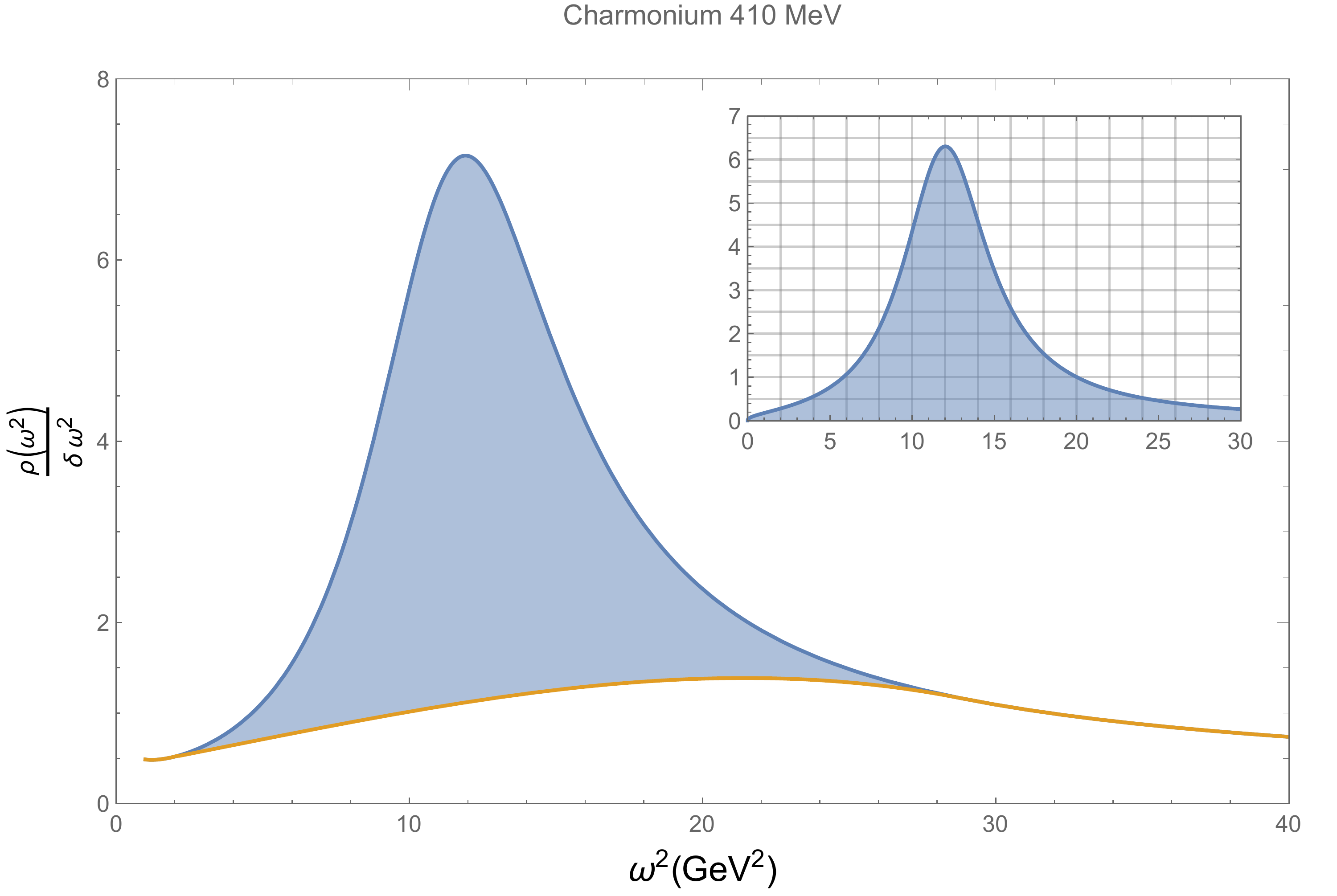}
    \includegraphics[width=3.4 in]{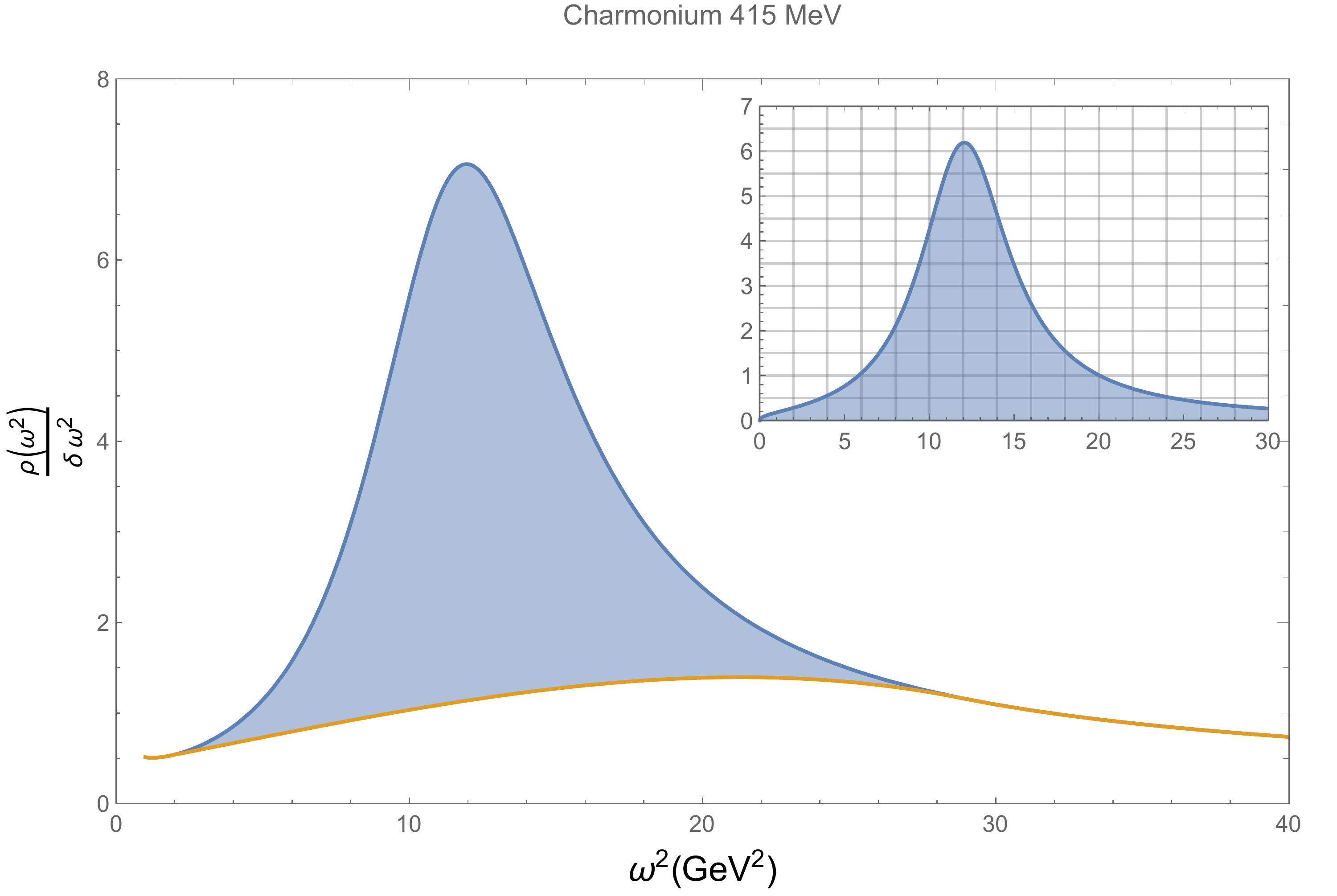}\\
    \includegraphics[width=3.4 in]{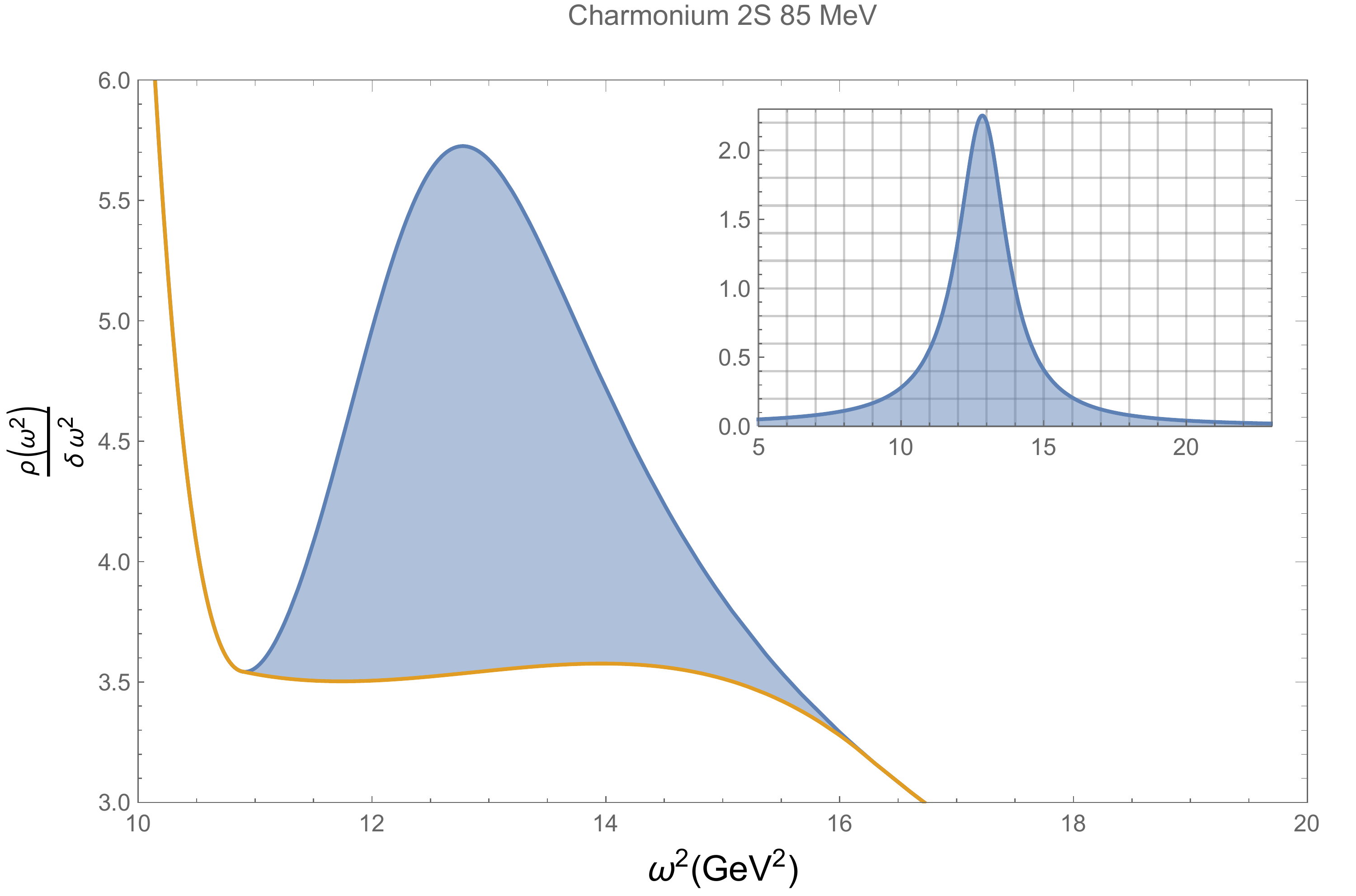}
    \includegraphics[width=3.4 in]{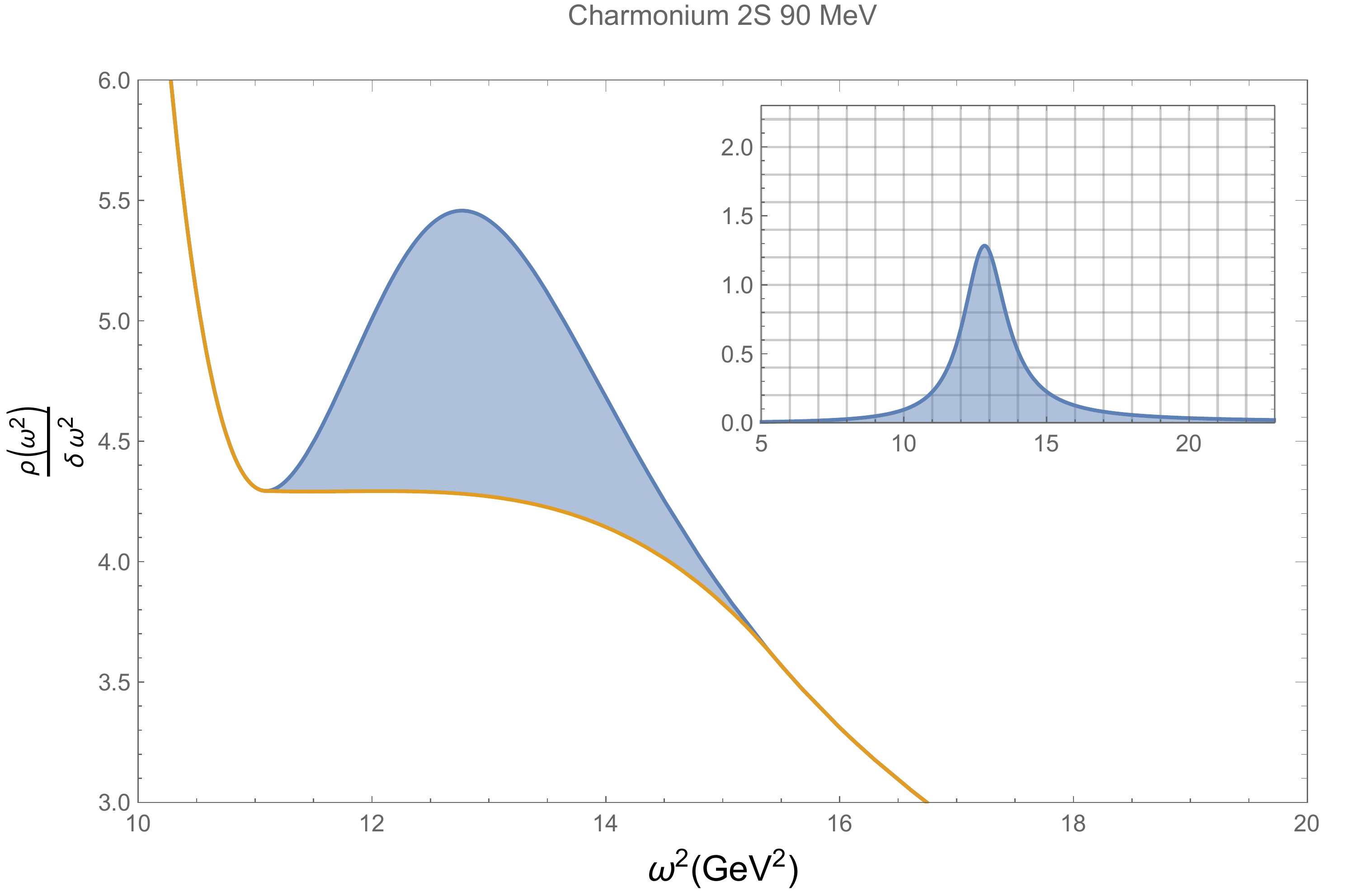}\\
     \includegraphics[width=3.4 in]{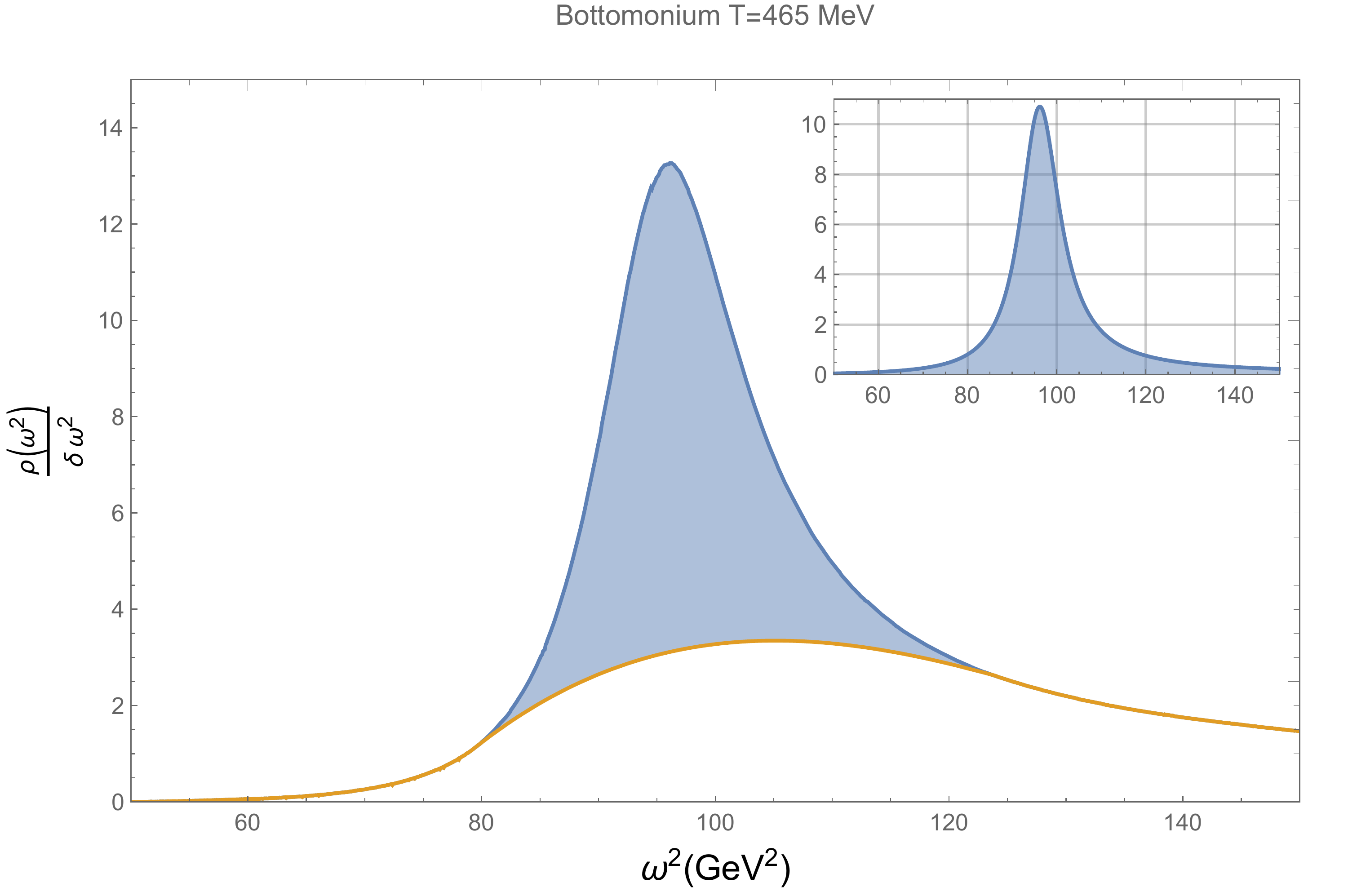}
    \includegraphics[width=3.4 in]{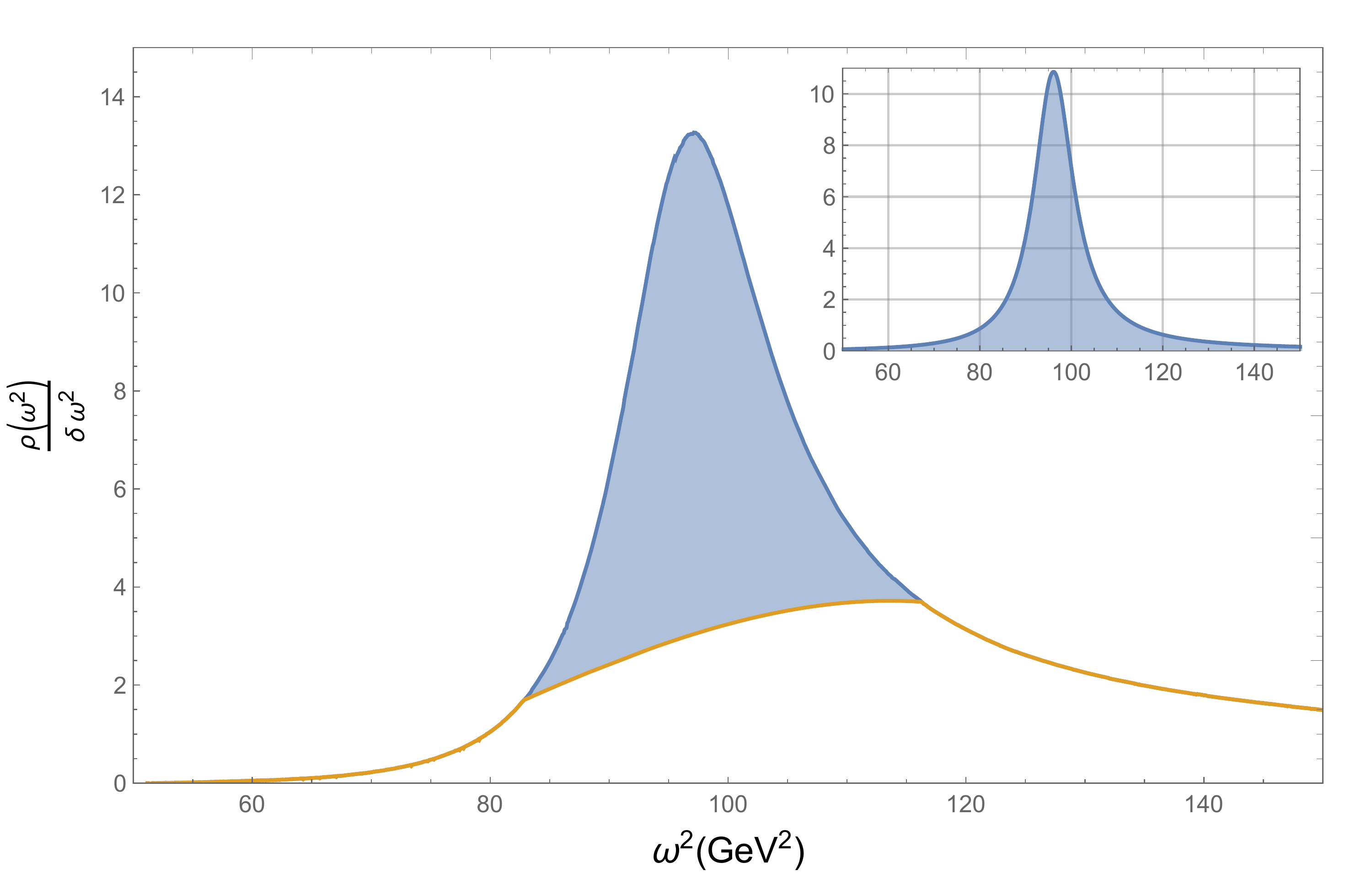}
  \end{tabular}

\caption{This figure depicts the subtraction procedure for $J/\psi$ at 400 MeV and 415 MeV, $\psi'$ at 85 MeV and 90 MeV, and $\Upsilon$ at 465 MeV. Notice that the background polynomial appears as the orange function in both cases. We plot the subtracted spectral density on the top right part of each figure that we fit with the Breit-Wigner distribution \eqref{BW-fit}. In the lower panels, we plot the bottomonium case for the same temperature, 465 MeV, with two different interpolating polynomials. In both situations, changing the polynomial does not affect the melting criterium. Recall that, unless other non-holographic effective models, the in-medium effects are encoded into the metric tensor. Thus, any proper characteristic behavior, as heavy quarkonium regeneration or Gluon radiation, is indistinguishable. }
\label{fig:three}
\end{figure*}
\end{center}

In the thermal approach to heavy quarkonium, the colored medium is vital since it strongly modifies the vacuum phenomenology. In particular, following the Feynman-Hellman theorem analysis, it is expected that bound states energy decrease when constituent mass is increased at zero temperature \cite{Quigg:1979vr}. Consequently, zero temperature spectral peaks experience shifting in their positions, color singlet excitations transform into other singlet states by thermal fluctuations, or these singlet excitations transform into another color octets. All of this intricated phenomenology is encoded in the medium. Therefore,  in order to isolate the thermal information regarding the heavy quarkonium state melting process,  a proper subtraction scheme is needed.  In our case, we will consider an interpolating polynomial in $\omega^2$ that will be subtracted to the spectral density, allowing us to obtain a Breit-Wigner distribution associated with the heavy quark state only. In figure \ref{fig:three}, we depict the subtraction process for the melting of $J/\psi$, observed in our model at 415 MeV (2.92 $T_c$).

At this step, an important remark should be made. The interpolating polynomial is not defined univocally. We can fix a criterium that these sorts of polynomial should obey. In principle, since we do not have a proper phenomenological tool at hand to split the behavior of the medium from the hadronic state, we will ask for a \emph{smooth subtraction}. In other words, the region where the interpolating polynomial splits from the spectral function should not display an abrupt change.  Since the possible functions that could match this condition are infinite, we can only bring a temperature interval where the meson starts to melt. However, choosing similar polynomials will lead to the same melting interval. See lower panels in figure \ref{fig:three}.

\subsection{Melting Temperature Criterium}
As we observe in figure \ref{fig:one}, mesonic states disappear progressively with increasing temperature. In the holographic potential case, the melting temperature is not connected with the disappearing of the confining barrier. Since the potential has a depth well in the UV region, the thermal stability would be associated with the tunneling of such a barrier. 

In the holographic situation, the generated dual object is a colored medium at thermal equilibrium, where the heavy quarkonium exists. In such a static situation, mesonic states either exist or have melted down. Thus, the only relevant information at the holographic level we have is the spectral function and the background subtraction.

In order to find the interval where heavy mesons start to melt, we will follow the standard criterium connecting the Breit-Wigner maximum with its graphical width, defined as a product of the meson mass and the thermal width

\begin{equation}
\frac{\bar{\rho}(\omega_0^2)}{\omega_0\,\frac{\Gamma}{2}} <1.   
\end{equation}

Notice that the definition depicted above is an alternative to the criteria defined from the effective potential models and lattice QCD, defined where the melting occurs when the in-medium binding energy equals the thermal decay width \cite{Rothkopf:2020vfz}.  In the holographic case, melting temperatures are intrinsically connected to decay constants, proportional to the two-point function residues at zero temperature.  Recall the decay constants carry information about how the mesonic states decay electromagnetically into leptons. Thus, indirectly they measure the mesonic stability affected by thermal changes: excited states with lower binding energy than the ground one melt first. This connection with meson stability is supported by the experimental fact that decay constants decrease with the excitation number.  Another possible form to explore the connection between the mesonic melting process and stability is done in the context of configurational entropy, discussed in refs. \cite{Braga:2018fyc,Braga:2018zlu,braga:2020myi,Braga:2020opg}. 

In the case of the charmonium, the $\psi'$ state melts near 90 MeV or $0.63\,T_c $. The ground state, the $J/\psi$ meson melts near to 415 MeV or $2.92 \,T_c$.  If we compare with the pNRQCD results \cite{Burnier:2015tda}, we obtain a lower temperature for the  $2\,S$ charmonium state (lattice result: $0.95\,T_c$) but higher for the ground state (lattice result: $1.37\,T_c$). The main difference in both results is that in our holographic case we are considering heavy quarkonium at rest, i.e., $|\vec{p}|=0$.

A similar situation is observed in the  bottomonium case: the $\Upsilon(2\,S)$ melts near to $115$ MeV (or $0.81\,T_c$), compared with the pNRQCD result of $1.25\,T_c$. For the ground state we have $465$ MeV (or $3.27\,T_c$), compared with the lattice result of $2.66\,T_c$. 

If we compare with holographic stringy models \cite{Andreev:2019hrk}, where the melting temperature is estimated from the string tension in an AdS deformed target space, we found bigger results for heavy quarkonium melting temperature. They predict $1.05\,T_c$ and $2.52\,T_c$ for charmonium and bottomonium respectively.

\subsection{Thermal Mass}
\begin{center}
\begin{figure*}
  \begin{tabular}{c c}
    \includegraphics[width=3.4 in]{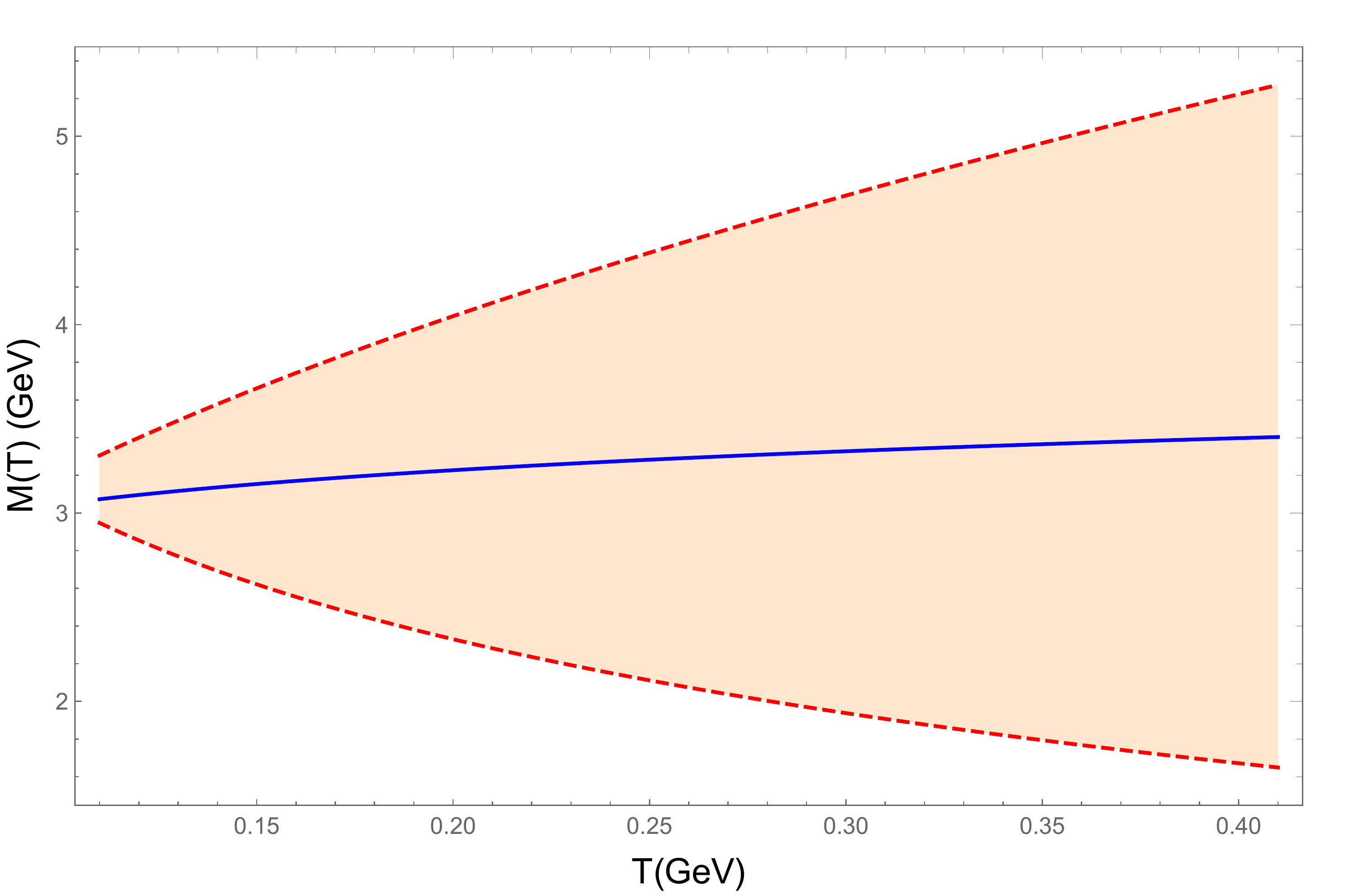}
    \includegraphics[width=3.4 in]{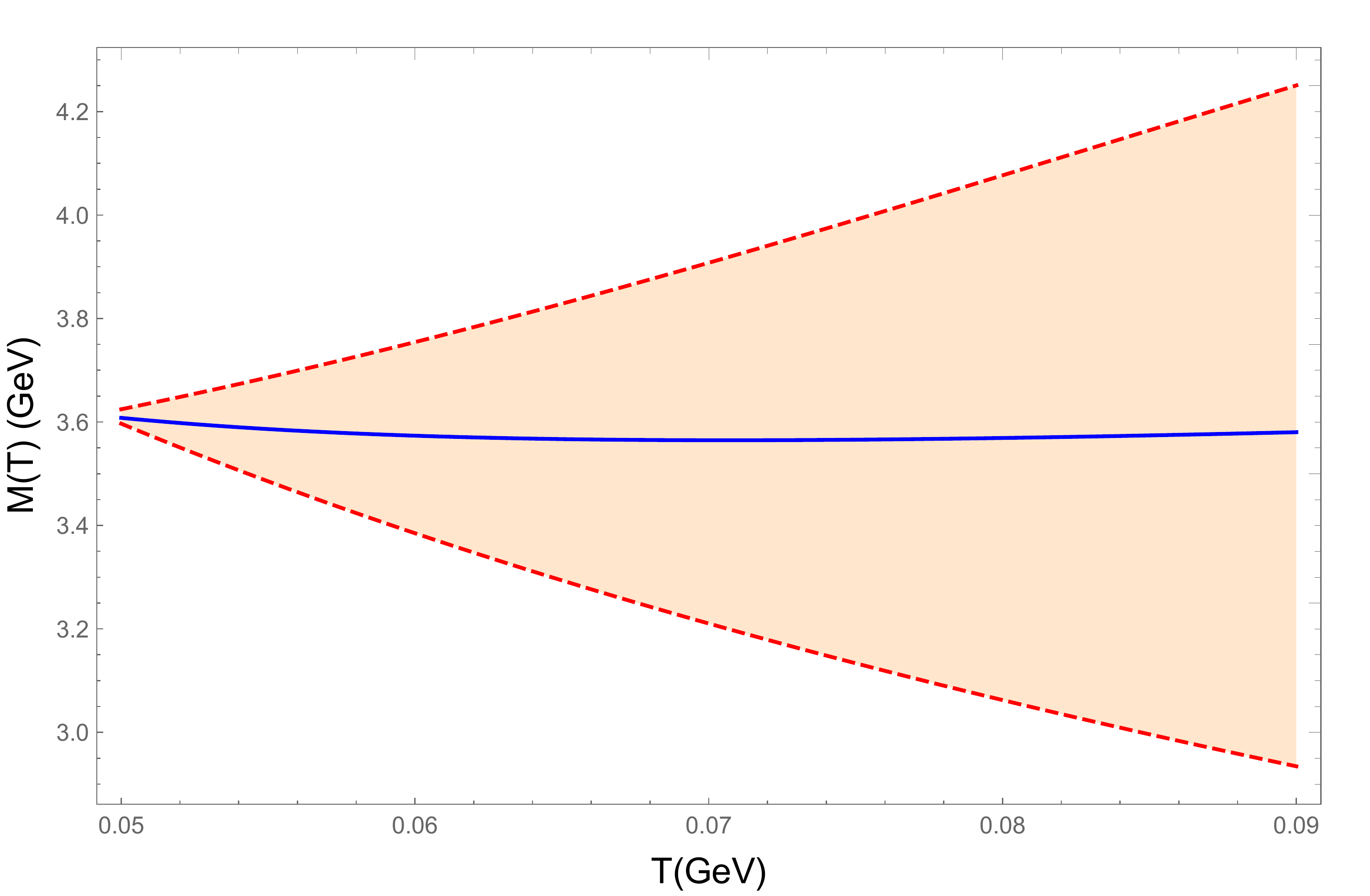}\\
    \includegraphics[width=3.4 in]{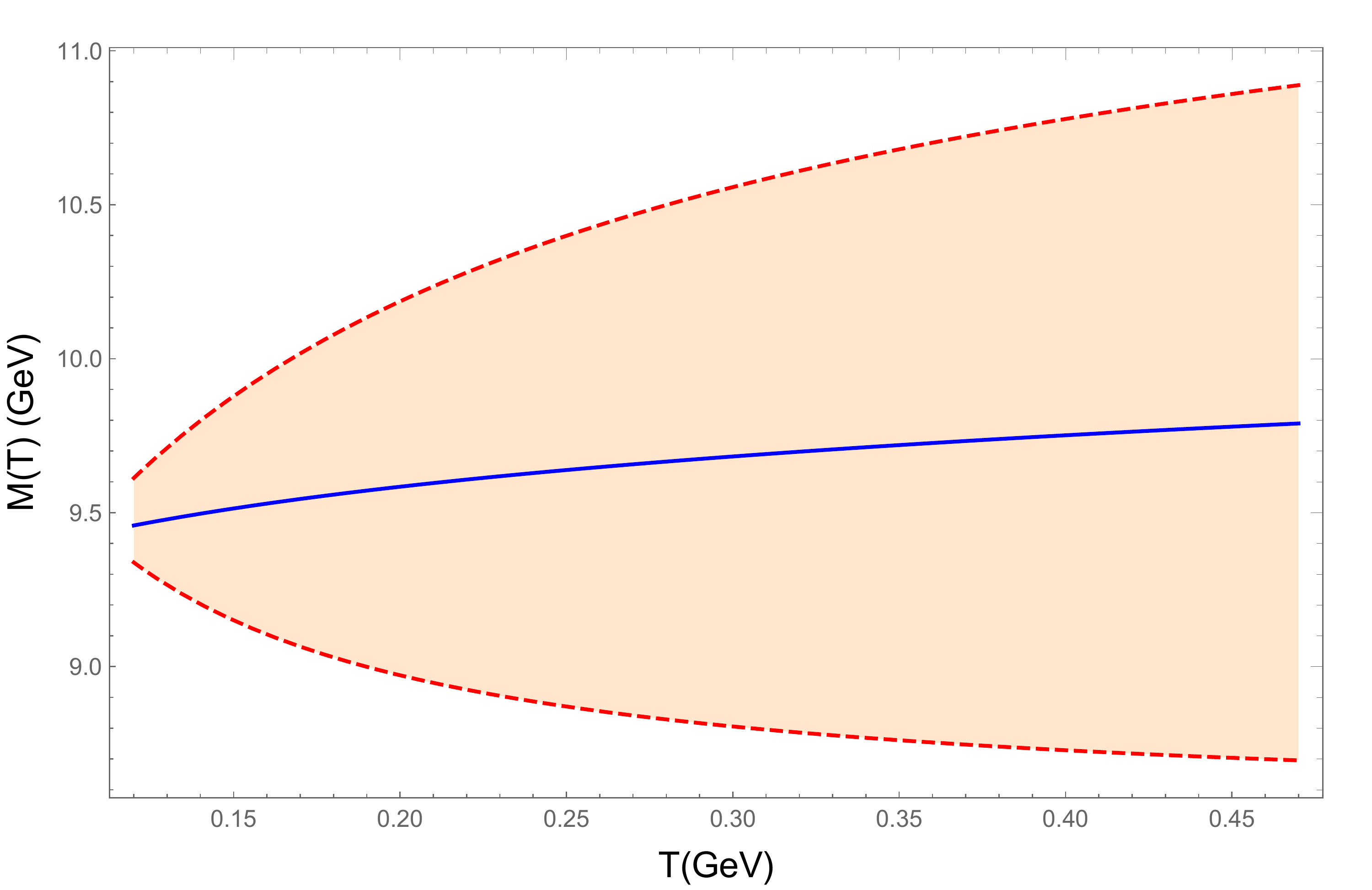}
    \includegraphics[width=3.4 in]{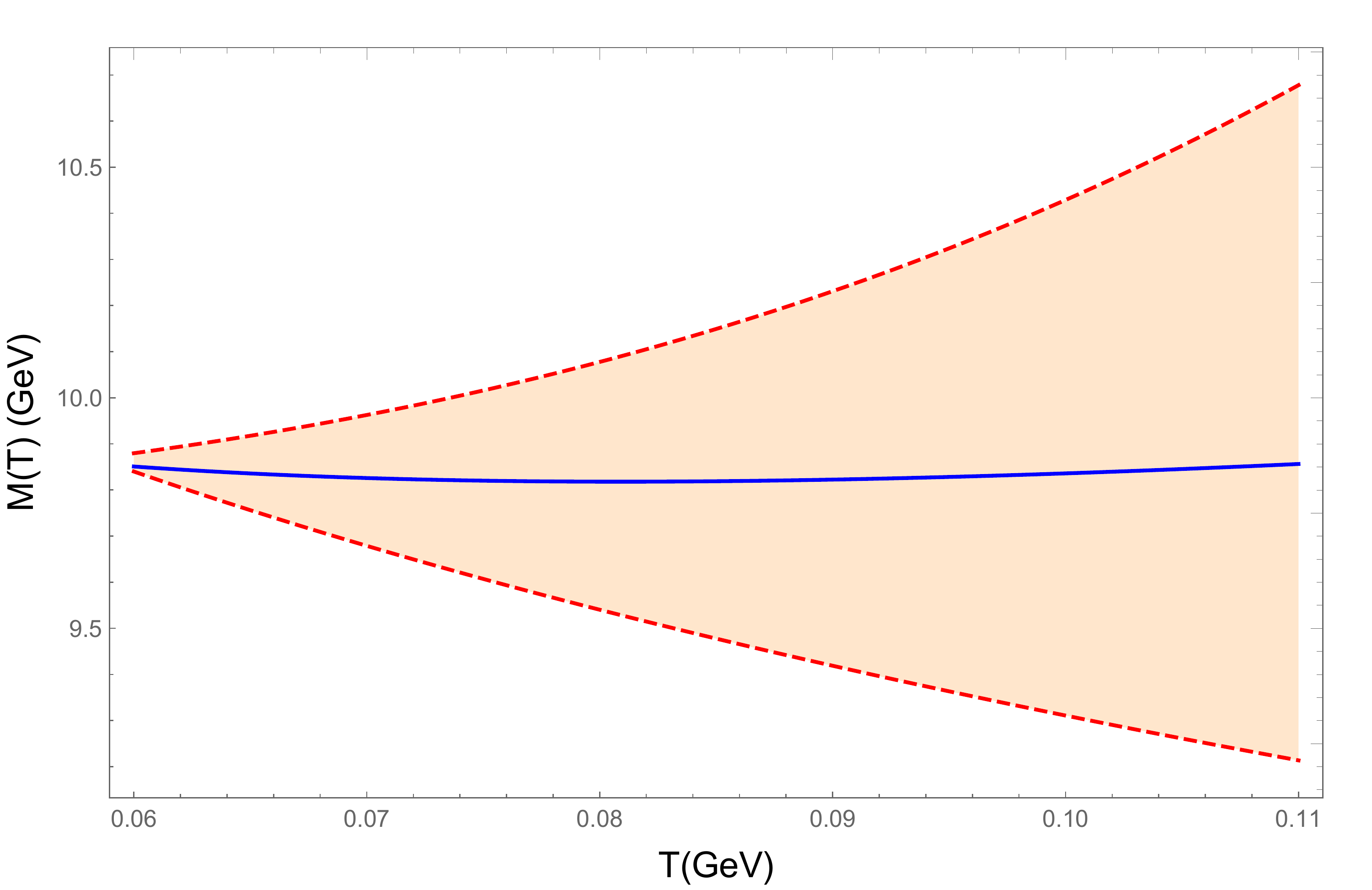}\\
  \end{tabular}

\caption{Resonance location as a function of the temperature. The shaded region in each panel describes the increase of the thermal width until the meson melting occurs. The left panels correspond to ground states, and the right panels  are the first excited states. Upper panels correspond to charmonium, and lower panels correspond to bottomonium. }
\label{fig:four}
\end{figure*}
\end{center}

Other important quantities to discuss are the masses and widths of the different hadronic states since these parameters have information about the interaction with the colored medium. Figure \ref{fig:four} has summarized the mass thermal behavior modeled for the first two charmonium and bottomonium excited states.  Comparing with other holographic models (see \cite{Fujita:2009ca,Fujita:2009wc} for heavy mesons; \cite{Colangelo:2009ra,Colangelo:2009ra} and \cite{Cui_2016} for light mesons), the mass for the ground state in our case tends to increase with temperature until the meson melting takes place, as the upper (J/$\psi$) and lower ($\Upsilon$) panels in figure \ref{fig:four} display. The same behavior is observed for the charmonium first excited state, depicted in Figure \ref{fig:four} right upper panel.  However, this very same behavior is not observed for the first excited state of the bottomonium. In the $\Upsilon(2S)$ meson case, the hadronic resonance location decreases with the temperature. 

The observed behavior for the thermal mass in our case seems to be quite different from the one depicted in \cite{Fujita:2009wc}. In their case, the thermal mass increases towards a maximum, where the authors claimed the melting process starts, and then thermal mass decreases up to the last charmonium meson is melted. In our case, such a concavity change occurs for low temperatures compared with $T_c$, far from the melting temperatures, around three times $T_c$.  The monotonicity of the thermal mass appears to be more consistent with lattice calculations \cite{Burnier:2016kqm,Rothkopf:2020vfz}. In those approaches, writing the NRQCD heavy quark potential is done in the soft scale, i.e., kinematical scale. In the case of hard scales, near to the constituent quark masses, other approaches are necessary.

In the context of QCD sum rules \cite{Dominguez:2009mk}, following the Hilbert moment mechanism, the thermal mass in the case of heavy quarks does not change with the temperature until the system reaches the critical temperature, where it drops. As an interesting observation, in this model, the decay constants go to zero as the temperature comes closer to the critical one, indicating that the melting has occurred. 
%%%%%%%%

\section{Conclusions}\label{con}
By deforming the non-quadratic dilaton defined in \cite{MartinContreras:2020cyg} using the proposal given by Braga et al. in \cite{Braga:2017bml}, it was possible to fit for the vector charmonium and bottomonium both the mass spectra as non-linear Regge trajectories and their decreasing decay constants. The precise holographic description of the heavy vector meson excited states is reached by considering all the lessons learned in the last decade of bottom-up AdS/QCD.

The precision of the fit is measured by the $\delta_{RMS}$, defined in eq.(\ref{delta}), being $6\%$ for charmonium and $7,2\%$ for bottomonium. The dilaton deformations are necessary for a precise description of the spectrum of masses and decay constants. If we use the original quadratic dilaton to describe the charmonium spectrum by fixing $k=1.55$ GeV, we find  $\delta_{RMS}=74\%$. So, the new parameters introduced in the dilaton do allow an accurate description of the spectrum.  Notice that the model has predictability even though we are using four parameters to fit each heavy quarkonium family. As a matter of fact, for the non-linear trajectory $M^2=a(n+b)^\nu$ we need three parameters. If we assume that decay constants are functions of the excitation number $n$ only, we can write them as $f(n)=c(-n+d)$, if we suppose linearity as our first guest. The minus sign in the parametrization emphasizes the decreasing behavior of the decays with $n$. Thus, if we count the maximum number of parameters need for both decays and masses, we obtain five parameters. If we assume non-linear behavior for decays, we have one extra parameter, implying six instead of five maximum parameters per family. Thus, in our case, we have four. Thus our model is predictable.
Such precision is essential to set the correct zero temperature behavior of the spectral functions. If we think of the increasing temperature as an analog for time evolution, zero-temperature properties play the role of initial conditions.

Spectral functions have been numerically computed for several representative values of the temperature. As expected, pronounced resonance peaks around the zero temperature masses of charmonium and bottomonium are observed near $T_c$. To discuss the fate of the particle states when increasing temperature, it is necessary to subtract background contributions from the spectral functions. We provide a detailed discussion on this subject and propose a numerical scheme to perform such a subtraction. The Breit-Wigner peaks are analyzed. We obtain the melting temperature of $J/\Psi$ and $\Upsilon$ to be, respectively, $T_{J/\Psi}=415$ MeV and $T_{\Upsilon}=465$ MeV. These high melting temperatures obtained are directly connected to the correct description of the decay constants of the corresponding fundamental states of $c\bar{c}$ and $b\bar{b}$. The excited states $\Psi',\Upsilon'$ melts at temperatures smaller them $T_c$.  So, we consider smaller temperatures around $50-60$ MeV where we can see the pronounced peaks associated with the states. Within this range of temperatures, around $50-470$ MeV, we consider the thermal mass shifting of $J/\Psi,\Psi'$ and $\Upsilon,\Upsilon'$. We observe a small and monotonic increase in the masses of the ground states with temperature.

The specific form of the dilaton leads to a holographic potential that differs from the one obtained in quadratic dilaton models. In the present case, there is a narrow well in the ultra-violet region. The melting of the fundamental state is no longer entirely governed by the disappearance of the infra-red barrier. For this shape of holographic potential, the criteria for defining the melting of the states established in  \cite{Vega:2018dgk} does not apply. It is a task for future work to understand the melting process from the thermal evolution of this class of holographic potentials.

\begin{acknowledgments}
We wish to acknowledge the financial support provided by FONDECYT (Chile) under Grants No. 1180753 (A. V.) and No. 3180592 (M. A. M. C.). Saulo Diles thanks the Campus Salinopolis of the Universidade Federal do Par\'a for the release of work hours for research.
\end{acknowledgments}
\bibliography{references}
\end{document}